\setbox1 = \hbox to 12pt { \vrule height 6pt \hfill \vrule height 6pt}
\setbox2= \vbox {  \hrule   \moveleft 1mm \box1 \hrule }

\def\carre{\hbox{\vrule \vbox to 7pt{\hrule width 6pt \vfill \hrule}\vrule }}

\def\R{{\rm I}\!{\rm R}}
\def\Z{{\rm Z}\!\!{\rm Z}}

\def\C{C^{\infty }}

\def\L{\Lambda }

\def\e{\varepsilon }

\def\@{\infty }
\def\&{\rightarrow }

\font\un = cmbx10 at 14pt \null \vskip 2cm \centerline {\un DECAY
OF QUANTUM CORRELATIONS IN A LATTICE }
\medskip
\centerline {\bf  BY HEAT KERNEL METHODS.} \vskip 1cm \centerline
{\bf L. AMOUR, C. CANCELIER, P. LEVY-BRUHL and J. NOURRIGAT}
\medskip
\centerline {D\'epartement de Math\'ematiques, UMR CNRS 6056}

\centerline {Universit\'e de Reims. B.P. 1039. 51687 Reims Cedex
2. France} \bigskip \centerline {Abstract}
\medskip
We prove some estimations of the correlation of two local
observables in quantum spin systems (with Schr\"odinger equations)
at large temperature. For that, we describe the heat kernel of the
Hamiltonian for a finite subset of the lattice, allowing the
dimension to tend to infinity. This is an improved version of an
earlier unpublished manuscript.

\bigskip

 \noindent {\bf 1. Introduction.}
\medskip
In the last years, many works were devoted to the estimates, or
asymptotics, of the correlation of two local observables, (or
Ursell functions of $n$ local  observables), for classical spin
systems, at large, or at small temperature. The aim of this work
is a beginning of a similar study  for quantum spin systems,
related to the Schr\"odinger equation, (for two observables, at
large temperature).
\medskip

 In classical statistical mechanics on a lattice $L$, an interaction
 is the definition, for each finite subset $\Lambda$ of $L$,
 of  a function
$V_{\Lambda}$ on $\R^{\Lambda}$, or $(\R^p)^{\Lambda}$ if there
are $p$ degrees of freedom at each site. Then, for each $\beta >0$
(the inverse of the temperature), the mean value of the local
observable $f$, (i. e. of a function $f$ on $(\R^p)^{\Lambda}$),
is:$$E_{\Lambda , \beta} (f) = Z_{\Lambda} ( \beta )^{-1} \int
_{(\R ^p)^{\Lambda}} e^{- \beta V_{\Lambda} (x) } f(x) dx \hskip
1cm Z_{\Lambda} ( \beta ) = \int _{(\R ^p)^{\Lambda}} e^{- \beta
V_{\Lambda} (x) } dx .$$ The bilinear analogue (correlation of two
local observables), or the multilinear analogue (Ursell function
of $n$ local observables), are defined in a standard way, (see,
for example, D. Ruelle [21], B. Simon [22] or R.A. Minlos [18]).
We say that $f$ is supported in a part $E$ of $\Lambda$ if $f$
depends only on the variables $x_{\lambda}$ corresponding to sites
$\lambda$ which are in $E$. A classical problem is to estimate the
decay of the correlation of two local observables $f$ and $g$ with
disjoint supports $E$ and $F$ when the distance of $E$ and $F$
tends to $+\infty$.  When $\beta $ is small enough (large
temperature), this is a classical result of L. Gross [10]. For the
study of Ursell functions at high temperature in classical
mechanics, (tree decay), see, for example, Bertini, Cirillo and
Oliveri [8]. For a small temperature, the problem is more
complicated and more hypotheses are needed: see the works of
Helffer, Sj\"ostrand, V. Bach, T. Jecko, J.S. M\"oller, O. Matte,
... [6],[7],  [11] to [17], [24], [25].

\bigskip
In this work, we are interested to similar results in quantum
statistical mechanics, for $\beta$ small enough.
Let us consider a quantum $d$-dimensional
lattice of particles, each of them moving in  ${\R}^{p}$. To each
finite subset $\Lambda $ of the lattice $L= \Z^d$, we shall define
below, with more details, a potential $V_{\Lambda}$, which is
still a real valued function on $(\R^p)^{\Lambda}$. We denote by
$H_{\Lambda }$ the following differential operator in $({\R}^{p
})^{\Lambda }$, depending on the Planck's constant $h$:

$$H_{\Lambda } \ = \ -{h^2\over 2} \ \sum _{\lambda \in \Lambda }
\Delta _{x_{\lambda }} \ +\ V_{\Lambda }(x) \leqno (1.1)$$ where
$x= (x_{\lambda })_{\lambda \in \Lambda }$ denotes the variable of
$({\R}^{p })^{\Lambda }$, each variable $x_{\lambda }$ being in
${\R}^{p}$. With suitable hypotheses (see below), the exponential
$e^{-\beta H_{\Lambda}}$ will be defined for $\beta >0$, and will
be of trace class. Then, a local observable is no more a function,
but a bounded operator $A$ on the Hilbert space ${\cal
H}_{\Lambda} = L^2 ((\R^p)^{\Lambda})$.
The mean value $E_{\Lambda, \beta } (A)$ is defined now,
instead of the previous definition, by:
$$E_{\Lambda , \beta } (A) = Z_{\Lambda} (\beta) ^{-1} \
 {\rm Tr}(e^{-\beta H_{\Lambda}}A),
\hskip 1.5cm Z_{\Lambda} (\beta) =  {\rm Tr}(e^{-\beta
H_{\Lambda}}). \leqno (1.2)$$
If $E$ is a subset of $\Lambda$, we
say that $A$ is supported in $E$ if $A$  can be seen as an
operator on the Hilbert space ${\cal H}_E$. If $A$ and $B$ are two
local observables, supported in two disjoint subsets $E_1$ and
$E_2$ of $\Lambda$, a natural definition for the correlation is:
$${\rm Cov }_{\Lambda , \beta} (A, B) = E_{\Lambda , \beta}(AB) -
E _{\Lambda , \beta} (A) E _{\Lambda , \beta} (B). \leqno (1.3) $$
One of the goals of this work is to give an analogue of the result
of L. Gross in this situation, and to estimate the decay of ${\rm
Cov }_{\Lambda , \beta} (A, B)$ when  the distance of  the
supports  $E_1$ and $E_2$ of $A$ and $B$ tends to $+\infty$ (see
Theorem 1.3 below), assuming that $\beta$ is small enough. Perhaps
such a bound can be obtained by probabilistic methods, (see [1],
[18], [19], $\ldots$), but we want here to prove it by a careful
study of the integral kernel of the operator $e^{-\beta
H_{\Lambda}}$.  Since  $| \Lambda|$ tens sometimes to $+ \infty $,
we have to study a heat kernel in large dimension, like in
Sj\"ostrand [23].
The bound of the correlations is applied to
study the rate of convergence to some thermodynamic limits, and to
prove that there is no  phase transitions (discontinuity of the
mean energy per site) for $\beta$ small enough (see Theorems 1.4
and 1.5 below).

\bigskip Now, let us give  more details on the {\it interaction},
i.e. on the family of functions  $(V_{\Lambda})_{\Lambda \subset
\Z^d} $. We consider a function $A \in C^{ \infty } ({\R}^{p }, \R
)$ and, for each pair of sites $\lambda$ and $\mu$ in the lattice,
a function $B_{\lambda , \mu }\in C^{ \infty } ({\R} ^{2p},\R )$.
For each finite subset $\Lambda $ of ${\Z}^d$, we denote by
$V_{\Lambda }$ the following potential in $({\R}^{p })^{\Lambda
}$: $$V_{\Lambda} (x)\ = \ \sum _{\lambda \in \Lambda } A
(x_{\lambda } )\ +\ \ \sum _{{\lambda , \mu \in\Lambda \atop
\lambda \not=\mu }} B_{\lambda , \mu }(x_{\lambda },x_{\mu })
\hskip 1cm x = (x_{\lambda })_{\lambda \in \Lambda }. \leqno
(1.4)$$ For sake of simplicity, we assume that $B_{\mu , \lambda}
= B_{\lambda , \mu}$. When invariance by translation is needed,
$B_{\lambda , \mu }$ will depend only on $\lambda - \mu$. We shall
assume that $B_{\lambda , \mu}$   is small when $|\lambda - \mu  |$
is large. More precisely, we assume that, for some $\varepsilon \in
]0, 1[$, the following hypothesis is satisfied:
\smallskip
\noindent $(H_{\varepsilon}$)  {\it For each $\alpha \geq 0$  and
$\beta \geq 0$,  there
 exists  $C_{\alpha , \beta }(\varepsilon)>0$  such  that
$$ \sup _{\lambda \in \Z^d} \sum _{\mu \in \Z^d} {\Vert
 \nabla _{x_{\lambda }}^{\alpha} \nabla _{x_{\mu}} ^{\beta
 }B_{\lambda , \mu }\Vert
 \over \varepsilon ^{|\lambda - \mu |} } \leq C_{\alpha
 , \beta } (\varepsilon).  \leqno (1.5)$$
 The function $A$ is  bounded from below. All the derivatives of order
 $\geq 1$ of $A $ are bounded.  For each  $m>0$ and $\beta >0$,
 we have:
 $$\sup _{ x\in \R^p} (1
+|x|)^m e^{ -\beta  A(x)} = C_m(\beta ) < + \infty .\leqno (1.6)$$}
\bigskip
In all this paper, we denote by $\Vert \ \Vert $ the $L^{\infty}$
norm of a function.  The hypothesis (1.6) is not needed for
Theorems 1.1 and 1.2, but only for Theorem 1.3, 1.4 and 1.5. For
the estimation of correlations (theorem 1.3), if the two
observables are multiplications by bounded functions, we don't
need (1.5) for all $\alpha$ and $\beta$, but for a finite number.
It is the same if the supports of the observables are, for
example, single points.
 If $(H_{\varepsilon})$ is
 satisfied for some $\varepsilon \in ]0, 1[$,
 there exist $M_1(\varepsilon)$ and $M_2(\varepsilon)$,
 (independent of $\Lambda$), such that,
  for each finite set $\Lambda$, and for each point $\lambda \in
 \Lambda$:
 $$ \sup _{\lambda \in \Lambda} \Vert \nabla _{x_{\lambda}} V_{\Lambda } \Vert \leq
 M_1(\varepsilon)
 \hskip 1cm
\sup _{\lambda \in \Lambda} \sum _{\mu \in \Lambda}
 { \Vert \nabla _{x_{\lambda}}
  \nabla _{x_{\mu}} V_{\Lambda } \Vert \over
  \varepsilon ^{ | \lambda - \mu |}}
  \leq  M_2(\varepsilon) .\leqno (1.7)$$
  \bigskip In the following, the parameter $\beta$ will be
  denoted by $t$, since we shall use evolution equations.
   All the results of this work, excepted
  the bounds for the first order derivatives,
  will be valid under the following
  condition:
  $$ ht = h\beta \leq T_0  \hskip 1cm T_0 =
  M_2(\varepsilon)^{-1/2}. \leqno (1.8)$$
  \bigskip
  The first result is devoted to the
description of the integral kernel of $e^{-tH_{\Lambda}}$, with
inequalities where the constants are independent of $\Lambda $, if
the condition   $(H_{\varepsilon})$ is satisfied. We denote  by
$\nabla _ {\lambda }$ the differential, with respect to
 $(x_{\lambda }, y_{\lambda })\in \R^{2p}$ ($\lambda \in \Lambda$)
 of a $\C $ function $f$ on $(\R^{2p})^{\Lambda }$. The norm of
 $\nabla _{\lambda}f(x)$ is its norm in $(\R^{2p})^{\star}$.
 We denote by ${\rm diam}(A)$ the diameter
 of a subset $A$ of $\Z^d$. (For the theorem 1.1, the norm in $\Z^d$
 is arbitrary.)

 \bigskip
\noindent {\bf Theorem 1.1.} {\it  Under the previous hypotheses,
the integral kernel $U_{\Lambda }(x, y, t )$ of $e^{-tH_{\Lambda
}}$ can be written  in the form $$U_{\Lambda }(x, y, t ) \ =\
(2\pi th^2)^{-p \vert \Lambda \vert /2} \ e^{-{ | x-y | ^2\over
2th^2}} \ e^{-\psi_{\Lambda }(x, y, t)}, \leqno (1.9)$$ where
$\psi_{\Lambda } $ is a $C^{\infty }$ function in $(\R^p)^{\Lambda
} \times (\R^p)^{\Lambda } \times [0, + \infty [$, depending on
the parameter $h>0$. Moreover, if $(H_{\varepsilon})$ is
satisfied, for each finite subset $\Lambda $ of $\Z ^d $,  we
have, for all $t>0$ $$\sup _{\lambda \in \Lambda}  \Vert \nabla
_{\lambda} \psi _{\Lambda}(., t) \Vert \leq t M_{1}( \varepsilon
).  \leqno (1.10)$$For each $m\geq 2$, for all points $\lambda
^{(1)}, ... , \lambda ^{(m-1)}$ in $\Lambda$,  we can write, if
$ht \leq T_0$, (the constant of (1.4)): $$\sup _{ (x, y) \in
(\R^p) ^{\Lambda} \times (\R^p) ^{\Lambda}} \sum _{\mu \in
\Lambda} {| \nabla _{\lambda ^{(1)}} ... \nabla _{\lambda
^{(m-1)}} \nabla _{\mu } \psi _{\Lambda} (x, y, t) | \over
\varepsilon ^{{\rm diam} (\{ \lambda ^{(1)}, ... \lambda ^{(m-1)},
\mu \} )}} \leq t K_{m}( \varepsilon) , \leqno (1.11)$$ where
$K_m(\varepsilon)$ is independent of $\Lambda$. }

\bigskip
J. Sj\"ostrand proved in [23] that, near the diagonal, the
integral kernel $U_{\Lambda }$ can be written in the form (1.9),
and he proved that an approximation modulo ${\cal O}(h^{\infty })$
of
 the function $\psi_{\Lambda } $ satisfies, near the diagonal,
 inequalities which are
  equivalent to (1.11). A family of functions satisfying the
  estimates (1.11) is called in [23] a
{\it  $0-$standard function with exponential weight}. Here, we
study the function $\psi_{\Lambda}$ itself, not an approximation,
and the estimation is global.
 \bigskip
In the semiclassical limit, $\psi_{\Lambda } (x, y, t)$ is
approximated by  the product of $t$ by the average of
$V_{\Lambda}$ on the segment $[x , y]$. By (1.4), this
semiclassical approximation is written as a sum of terms,
associated to points $\lambda$, or couples of points $(\lambda ,
\mu)$.  In the next theorem, we shall describe $\psi_{\Lambda } $
itself, and not its approximation, in a similar way. In fact,
there will be some difference: instead of a sum taken on the sites
$\lambda$ or the couples of sites $(\lambda , \mu)$ like in (1.4),
we shall need, for $\psi _{\Lambda}$, a sum of functions $T_Q \psi
_{\Lambda }$, associated to all the boxes $Q$ contained in
$\Lambda$. When it is restricted to the diagonal, the function
associated to the box $Q$ will depend only on the variables
$x_{\lambda}$ ($\lambda \in Q$), and the function will decrease
almost like $\varepsilon^{{\rm diam }(Q)}$ when $Q$ is large. In
the literature on classical lattice spin systems, (without
Schr\"odinger equation), the potential $V_{\Lambda }$ is often
supposed to be  a sum of such functions. A {\it box} of $\Z^d$  is
a set of the following form: $$\Lambda = \prod _{j=1}^d [a_j ,
b_j] \leqno (1.12)$$ where $a_j$ and $b_j$ are in $\Z$ ($a_j \leq
b_j$). In Theorem 1.1, the choice of the norm in $\Z^d$ was
irrelevant,
 but now it is the $\ell ^{\infty }$ norm.
\bigskip
\noindent {\bf Theorem 1.2.} {\it We can define, for each box
$\Lambda$ of $\Z^d$, and for each box
 $Q \subseteq \Lambda$, ($Q$ may be a single point),
 a function $(T_Q \psi _{\Lambda }) (x, y , t)$
 (where $\psi _{\Lambda}$ is the function of Theorem 1.1), such
 that:
 \smallskip \noindent 1. The function $(T_Q \psi _{\Lambda }) (x, y , t)$
 is $\C$ and depends only on the variables $x_{\lambda}$
 and $y_{\lambda}$ such that $\lambda \in Q$, and on the variables
 $x_{\lambda}- y_{\lambda}$ such that $\lambda \notin Q$, (restricted
 to the diagonal, this function is supported in $Q$).

 \smallskip \noindent 2.  We have:
 $$\psi _{\Lambda } (x, y, t) = \psi _{\Lambda } (0, y-x, t) +
 \sum _{Q \subseteq \Lambda } (T_Q \psi _{\Lambda }) (x, y , t)
 \leqno (1.13)$$
 where the the sum is taken on all boxes $Q$ contained in
 $\Lambda$, including the points.

\smallskip \noindent 3.  If $(H_{\varepsilon})$ is
satisfied ($0 < \varepsilon <1$),  if $ht \leq T_0$ (defined in
(1.8) and (1.7)), for each integer $m\geq 1$,  for each points
 $\lambda ^{(1)}$, $\ldots$, $\lambda ^{(m)}$ in $\Lambda $,
 we have, for some constant
 $K_m(\varepsilon )>0$ independent of $\Lambda$:
 $$\Vert \nabla _{ \lambda ^{(1)}} \ldots \nabla _{ \lambda ^{(m)}}
\big (T_Q \psi _{\Lambda } \big ) (., ., t, h) \Vert \leq
K_m(\varepsilon ) t \varepsilon ^{ {\rm diam }(Q \cup \{ \lambda
^{(1)} , \ldots, \lambda ^{(m)} \} )} (1 + {\rm diam} (Q))^{2d}.
\leqno (1.14) $$ If $m=0$, this result is valid for boxes $Q$ not
reduced to single points. If $m=0$ and $Q$ is a single point
$\lambda$, we can write: $$ \Big | T_{ \{ \lambda \} } \psi
_{\Lambda } (x, y, t) - t \widetilde A (x_{\lambda }, y_{\lambda})
+ t \widetilde A (0,  y_{\lambda} - x_{\lambda }) \Big |\leq
K(\varepsilon ) ( t + h^2 t^2 ). \leqno (1.15)$$ where $A$ is the
function appearing in (1.4), and: $$\widetilde A (x_{\lambda },
y_{\lambda}) = \int _0^1 A (y_{ \lambda} + \theta (x_{\lambda } -
y_{ \lambda} )) d \theta . \leqno (1.16)$$

}
\bigskip

Now, we shall apply the decomposition of Theorem 1.2 to the study
of correlations. For each disjoint subsets $E_1$ and $E_2$ of
$\Z^d$, let us denote by $K_{ \rm op , op }(E_1 , E_2 , t , h)$
the smallest positive constant such that, for each $A\in {\cal L}
( {\cal H} _{E_1})$ and $B\in {\cal L} ( {\cal H} _{E_2})$, for
each box $\Lambda$ containing $E_1$ and $E_2$, we have: $$| {\rm
Cov} _{\Lambda , t} (A , B) | \leq \ K_{\rm op , op } (E_1 , E_2 ,
t , h) \ \Vert A \Vert \ \Vert B \Vert .   \leqno (1.17)$$ By  the
definitions (1.2) and (1.3), this constant exists, and is $\leq
2$. Let $K_{ \rm fc , fc}(E_1 , E_2 , t , h)$ the smallest
positive constant such that, for each continuous, bounded
functions $f$ and $g$, supported in $E_1$ and $E_2$,  for each box
$\Lambda$ containing $E_1$ and $E_2$, we have: $$| {\rm Cov}
_{\Lambda , t} (M_f , M_g) | \leq \ K_{ \rm fc , fc}(E_1 , E_2 , t
, h) \  \Vert f \Vert \ \Vert g \Vert . \leqno (1.18)$$We define
similarly $K_{ \rm fc , op}(E_1 , E_2 , t , h)$ and $K_{ \rm op ,
fc}(E_1 , E_2 , t , h)$.

\bigskip
\noindent {\bf Theorem 1.3.} { \it  Let $(V_{\Lambda} ) _{\Lambda
\subset \Z^d}$ be an interaction satisfying $(H_{\varepsilon})$,
($0<\varepsilon < 1$). Then, for each $\delta $ such that $0 <
\varepsilon < \delta < 1$, there exists $t_1 (\varepsilon,
\delta)$ and functions $M(N, t, h , \varepsilon, \delta)$ and
$N(\varepsilon , \delta)$ with the following property. If  $ht
\leq T_0$ (defined in (1.8)), and if $t \leq t_1 (\varepsilon ,
\delta)$, for each finite disjoint sets $E_1$ and $E_2$, we have:
 $$K_{ \rm op , op }(E_1 ,
E_2 , t , h) \leq M( | E_1 \cup E_2 | , t, h, \varepsilon , \delta
)\ \delta ^{{\rm dist}(E_1 , E_2)}. \leqno a)$$ $$K_{ \rm fc , fc
}(E_1 , E_2 , t , h) \leq  \ t\ \inf \Big (|E_1| , |E_2| \Big ) \
N(\varepsilon , \delta )\ \ \delta ^{{\rm dist} (E_1 , E_2)}.
\leqno b)$$
 $$K_{ \rm op , fc }(E_1 , E_2 , t , h)
\leq  \ M( | E_1 | , t, h, \varepsilon , \delta ) \ \delta ^{{\rm
dist} (E_1 , E_2)}. \leqno c)$$

\smallskip \noindent If $f$ has bounded derivatives, and $g$ is
bounded, we can write: $$|{\rm Cov}_{\Lambda , t} (M_f , M_g) |
\leq \ \Vert \nabla f \Vert _{\infty} \ \Vert g\Vert \ M( |E_1|
,t, h, \varepsilon , \delta ) \ \delta ^{{\rm dist} (E_1 , E_2)}.
\leqno d)$$ The functions denoted by $M$, as functions of $t$ and
$h$, for fixed $\varepsilon$ and $\delta$, are bounded on each
compact of the set $\{ (t, h) , \ h>0 , 0 < t < t_1(\varepsilon,
\delta), ht < T_0 \}$. The constant $t_1(\varepsilon, \delta )$,
limiting the validity of the result, is independent of the three
sets $E_1$, $E_2$ and $\Lambda$.
 }
\bigskip
We don't give precisely the behaviour of the constants as
functions of $t= \beta$ when $t\rightarrow 0$, excepted in the
case of multiplications by bounded functions. It depends of the
behaviour of $C_m(t)$ in the hypothesis $(H_{\varepsilon})$.  For
the proof of this theorem, we obtain, from Theorem 1.2, a
decomposition of the integral kernel of $e^{-\beta H_{\Lambda}}$
which is similar to the Mayer decomposition in classical
statistical mechanics (see B. Simon [22]).

\bigskip
Now, we  are interested to thermodynamic limits, and to the rate
of convergence to such limits. We say that a local observable $A
\in {\cal L} ( {\cal H}_{\Lambda})$ is supported in a subset
$Q\subset \Lambda$ if $A$ can be seen also as an element of ${\cal
L} ( {\cal H}_Q)$. The proof of the next theorems 1.4 and 1.5
relies on  the estimations of correlations, (Theorem 1.3), and
also, directly  on the decomposition of $\psi_{\Lambda}$ (Theorem
1.2).

 \bigskip
 \noindent {\bf Theorem 1.4.} {\it
 If the interaction satisfies $(H_{\varepsilon})$,  if
 $ht \leq T_0$ (defined in (1.8)), and if $t$ is small enough, for each
 local observable $A $,  the following
thermodynamic limit exists: $$\omega _{t}(A) = \lim _{n\rightarrow
+ \infty} E_{\Lambda _n , t} (A) \hskip 1cm \Lambda _n = \{ -n,
\ldots , n \} ^d .\leqno(1.19)$$ Moreover, if $\varepsilon <
\delta < 1$, there exists $t_1 (\varepsilon, \delta)$ and a
function $K(h, t, \varepsilon, \delta ,N )$ such that, if $ht \leq
T_0$, if $t\leq  t_1(\varepsilon , \delta)$ and if $\Lambda _n$
contains the support of $A$, $$\Big | \omega _{t}(A)  - E_{\Lambda
_n , t}  (A)\Big | \leq K(h, t, \varepsilon, \delta ,|  {\rm supp}
(A)|  )\  \Vert A \Vert \  \delta ^{ {\rm dist} ({\rm supp} (A) ,
\Lambda _n^{c}) } . \leqno  (1.20)$$ }

\bigskip
Theorems 1.3 and 1.4 can be applied to prove some properties of a
state of the $C^{\star}$ algebra ${\cal A}$ associated to the
family of Hilbert spaces ${\cal H}_{\Lambda}$. Let us recall (see
B. Simon [22], section II.1, or Bratteli-Robinson [9]), that, if
$\Lambda _1 \subset \Lambda _2$, we have a natural identification
of ${\cal L}({\cal H} _{\Lambda _1})$ as a subspace of ${\cal
L}({\cal H} _{\Lambda _2})$, and ${\cal A}$ is the closure of the
union of (equivalence classes of) all the ${\cal L}({\cal H}
_{\Lambda })$. Then, for each $h$ and $t$ such that Theorem 1.4
can be applied, the limit in (1.19) defines a state on ${\cal A}$,
still denoted by $\omega _t$. Theorem 1.3 proves that this state
has the  {\it mixing property}, (quantum analogue of the
definition III.1.21 of B. Simon [22]): $$ \lim _{|u | \rightarrow
\infty } \Big [ \omega _t ( A \circ \tau _u B) - \omega _t (A)
\omega_t (B) \Big ] = 0     \ \ \ \ \ \ \ \forall A , B \in {\cal
A} \leqno (1.21)$$ if $ht < T_0$ and if $t$ is small enough. Here
$\tau _u$ is the natural translation by a vector $u\in \Z^d$. For
this application, it is useful that the condition of validity of
Theorem 1.3 does not depend on the number of elements of the
supports. In this application to the mixing property, invariance
by translation is needed, and we assume that $B_{\lambda , \mu }$
depends only on $\lambda - \mu$.

\bigskip
In the second application, we consider the mean value, not
of a local observable, but of the mean energy per site.

 \bigskip
 \noindent {\bf Theorem 1.5.}  {\it If $(H_{\varepsilon})$ is satisfied,
 and if $ht \leq T_0$, the following limit exists:
 $$ E(t) = \lim _{n\rightarrow + \infty} {1 \over  | \Lambda  _n | }
  X _{ \Lambda _n} (t), \ \ \ \ \   X _{ \Lambda } (t)  =   {\partial \over
  \partial t} \ln Z_{\Lambda } (t),
  \leqno (1.22)$$
 where $Z_{\Lambda}(t)$ is defined in (1.2) and $\Lambda _n$ in (1.19).
 We can write :
  $$\Big |   E(t) -    {1 \over  | \Lambda  _n | }
  X _{ \Lambda _n} (t)   \Big |  \leq { K(t, h) \over n} \leqno (1.23)$$
  The constant $K(t, h)$ is bounded on each compact set of
  $\{ ( h, t) , \ h>0, t>0, \ \ ht < T_0\}$.
 }
\bigskip
By the last statement, $E(t)$ is a continuous function of $t$ in
the domain in which the theorem is applicable: in other words,
there is no phase transition, if $t= \beta$ is small enough.

\bigskip
Theorem 1.1 is proved in section 2.
 The family of operators $T_Q$, which gives a decomposition
 of any function on $(\R^p)^{\Lambda}$, and that we use for the proof
 of theorem 1.2, is explained in
section 3.  The estimations of $T_Q \psi _{\Lambda}$ are proved in
section 4. In sections 5 and 6, we see how to dissociate two
disjoint sets $\Lambda _1$ and $\Lambda _2$ when $\Lambda $ is
their union. Then, in sections 7 and 8, the decomposition of
Theorem 1.2 will be applied to the correlations.  For the bound of
correlations in the multiplicative case, section 8 is not needed.
Section 9 is devoted to the proof of Theorems 1.4 and 1.5.
\bigskip
This article is an improved version of an earlier unpublished
manuscript [3]. We are very grateful to B. Helffer, T. Jecko, J.
S. M\"oller, V. Tchoulaevski and  V. Zagrebnov for helpful
discussions.

\bigskip
\noindent {\bf 2. Proof of Theorem 1.1. }
\bigskip
The heat kernel $U_{\Lambda}(x, y, t)$ must satisfy ${\partial U
_{\Lambda } \over \partial t}\ - {h^2 \over 2} \Delta _x
U_{\Lambda}  + V_{\Lambda} (x) U_{\Lambda} = 0$ for $t>0$.
Therefore, if $U_{\Lambda}$ is written in the form (1.9), the
function $\psi _{\Lambda }$ in $(\R^p)^{\Lambda}$ appearing in
this expression must satisfy
 the Cauchy problem:
 $${\partial \psi _{\Lambda } \over \partial t}\ + \
{x-y \over t}\ .\ \  \nabla _x \psi _{\Lambda } - {h^2 \over 2 }
\Delta _x \psi _{\Lambda } \ =\ V_{\Lambda }(x) \ -\ {h^2 \over 2}
\vert \nabla _x \psi _{\Lambda } \vert ^2 \leqno (2.1)$$ $$\psi
_{\Lambda }(x, y, 0, h )= 0\leqno (2.2)$$
\bigskip
This section is devoted to the study of this Cauchy problem. We
shall use a   maximum principle  for the linearized of (2.1), and
more generally, for operators $L_a$ in $(\R^p)^{\Lambda } \times
[0, T]$ $( \Lambda \in \Z^d , T>0)$ of the following form: $$ \Big
( L_au \Big ) = {\partial u \over
\partial t } \ +\ {x-y \over t} . \nabla u \ - \ {h^2 \over 2}
\Delta u \ + \ \sum _{\mu \in \Lambda } ( a_{\mu } (x, t)\ .\
\nabla _{x_{\mu }} u ) , \leqno (2.3)$$
where $a = (a_{\lambda})_{\lambda
\in \Lambda}$ is a family of  continuous and bounded
functions in $(\R^p)^{\Lambda }\times [0, T]$, and
$ y \in (\R^p)^{\Lambda }$. Since there is a drift
with unbounded coefficients, and since this maximum principle
will be used again, it was perhaps useful to give a precise statement:

\bigskip
\noindent {\bf Proposition 2.1.}{ \it Let $y\in (\R^p)^{\Lambda
}$, let $a = (a_{\lambda })_{\lambda \in \Lambda}$ be a family of
functions such that  $a_{\lambda }(x, t)$ $(\lambda \in \Lambda)$
is continuous and bounded  in $(\R^p)^{\Lambda } \times [0, T]$
$(T>0)$, with values in $\R^p$, and $u$ be a function in $C(
(\R^p)^{\Lambda } \times [0, T]) \bigcap C^2 ((\R^p)^{\Lambda }
\times ]0, T])$ such that $u$ and $\nabla _{x_{\lambda }} u$
$(\lambda \in \Lambda )$ are bounded in $(\R^p)^{\Lambda } \times
] 0, T]$ and $u(x,0)=0$. Assume that the function
 $f = L_a u$, defined by (2.3), (where $h>0$), is bounded.
  Then we have,
for each $t_0$ and $t$ ($0\leq t_0 \leq t \leq T$), $$\Vert u(.\ ,
t)\Vert  \ \leq \ \Vert u(.\ , t_0)\Vert + \int _{t_0}^t \Vert
L_au (., \ s)\Vert  \ ds .  \leqno (2.4)$$ }
\bigskip
\noindent {\it Proof.} Let $\chi\in C^\infty((\R^p)^\L)$ be a
real-valued function with $\chi(x)=1$ if $\vert x\vert\leq 1$ and
$\chi(x)=0$ if $\vert x\vert\geq 2$. For $R\geq 0$, set
$\chi_R(x,y,t)=\chi({\vert x-y\vert\over Rt})$ , $x\in (\R^p)^\L$,
$t>0$. If $0 < t_0 < t$, $y$ fixed, and $R>0$, the standard
maximum principle, applied in $(\R^p)^{\Lambda} \times [t_0 , t]$
with bounded coefficients in the first order terms, gives $$\Vert
\chi_R u (. , t) \Vert \leq \Vert \chi_R u (. , t_0) \Vert + \int
_{t_0}^t \Vert L_a ( \chi_R u (. , s) \Vert ds.$$ An explicit
computation of $ L_a ( \chi_R u (. , s)$ with our cut-off function
shows that, when $R \rightarrow + \infty $, $ L_a ( \chi_R u (. ,
s)\rightarrow L_a u$. The Proposition follows (we may also let
$t_0 \rightarrow 0$).

\bigskip
Now, we give a result of global existence of the solution $\psi
_{\Lambda}$ of the Cauchy problem (2.1), (2.2), with global bounds
of all the derivatives of this function, but, at this step, all
the bounds, excepted for the first order derivatives, may still
depend on $\Lambda$, hence on the number of sites in the lattice,
which will tend to infinity later.
 \bigskip
 \noindent
{\bf Proposition 2.2. } {\it Assuming only that $V_{\Lambda}$ is
in $\C ((\R^p)^{\Lambda }) $, real-valued, and that all its
derivatives of order $\geq 1$ are bounded, then, there exists a
unique global classical solution $\psi _{\Lambda } (x,y, t)$ in
$\C ((\R^p)^{\Lambda } \times (\R^p)^{\Lambda } \times [0, +
\infty [)$ to (2.1), (2.2). Moreover,
we have the estimations (1.10)
for the first order derivatives, with $M_1(\varepsilon)$ defined
in (1.7).  For each  $T>0$ and
$h>0$, all the derivatives of order $\geq 2$ with respect to $x$ and
$y$ of $\psi _{\Lambda }$ are bounded on $(\R^p)^{\Lambda } \times
(\R^p)^{\Lambda } \times [0, T]$ (with bounds which may depend, at
this step, on $\Lambda $, $T$ and $h$). }
 \bigskip
 \noindent {\it Proof}.
This result of existence is an adaptation, due to the presence of
the term containing ${x-y \over t}$, of a similar result about a
non linear heat equation, proved in [2]. Let us only explain what
is new here. In order to solve (2.1), (2.2)  by a fixed point theorem,
we use the explicit solution of the following Cauchy
problem $${\partial u \over \partial t}\ + \ {x-y \over t} \nabla
_x u - {h^2 \over 2 } \Delta _x u \ = \ f(x, t) \ \ \   (t>t_0), \hskip
1cm u (x, y, t_0)= g(x), \leqno (2.5)$$ which is given by
$$u (x, y, t)\ =\ \int _{  (\R^p) ^{\Lambda} } G_h(x, x', y, t_0, t)
g(x')dx' \ + \ \int _{ x'\in (\R^p) ^{\Lambda}  \atop
s \in [t_0 , t] }
G_h(x, x', y, s, t) \ f(x', s)\ dsdx' $$ where we set, if
$0<s<t$:
$$G_h(x, x', y, s, t)\ =\
\left ( { a(s, t) \over 2 \pi h^2 } \right )^{ {p\over 2}  |\Lambda|}
e^{ -{a(s, t) \over 2 h^2}
 | x' - m(x, y, s, t) | ^2} \leqno (2.6)$$
$$a(s, t)\ =\ {t\over  s(t-s)} \hskip 1cm m(x, y, s, t)\ =\ \left
( 1 - {s\over t} \right ) y \ +\ {s\over t} x . $$
We shall use the following properties of this kernel, where
$C>0$ is independent of all the parameters:
 $$ \int  _{ (\R^p) ^{\Lambda} }| \nabla _{x_{\lambda}}  G_h(x, x', y, s, t)  |  dx'
 \ \leq \
 {C\over h}\ \sqrt {{s\over t(t-s)}}$$
 To simplify the notations, we assume here that $p=1$, and we
  denote by ${\bf G}_h (y, s, t)$ (resp. ${\bf G} _h^{ (\lambda)} (y, s, t)$)
  the operator with integral kernel
  $(x, x')\rightarrow G_h(x, x', y, s, t)$
  (resp. $\partial _{x_{\lambda}}G_h(x, x', y, s, t)$). With these
  notations, we have:
  $$ \partial _{x_{\lambda}} \left (
{\bf G}_ h (y, s, t)f\right ) \ =\ {s  \over t}\  {\bf G}_h (y, s,t)\
\partial x_{\lambda} f. $$
Using this  operator, we study the integral equation satisfied,
not by $\psi _{\Lambda}$
itself, but by its  derivatives, if we want to find
$\psi _{\Lambda}$ satisfying (2.1) and (2.2).
For example, if $\psi _{\Lambda}$ satisfies (2.1) in an interval $[t_0, t_1]$,
if $y$ is fixed, and if we are given $\varphi _{y, \lambda} = \nabla
_{x_{\lambda}} \psi _{\Lambda} (. , y, t_0)$,
we hope that
the  first order derivatives $u_{y, \lambda} (x, t) =
\nabla_{x_{\lambda}}\psi_{\Lambda} (x, y, t)$ will satisfy,
setting $u_y= (u_{y, \lambda} )_{\lambda \in \Lambda}$ and
$\varphi _y = (\varphi _{y, \lambda} )_{\lambda \in \Lambda}$:
$$ u_y= S_y \varphi _y +  T_y (u_y) \hskip 1cm \big (
S_y \varphi  \big )_{\lambda}(., t)={t_0\over t} {\bf G} _h (y, t_0, t)
\varphi _{\lambda} , $$
$$\big ( T_y(u) \big )_{\lambda}(., t) =
\int _{t_0}^t {s\over t} {\bf G}_h (y, s, t) (\nabla _{x_{\lambda}}
V_{\Lambda} ) ds - {h^2 \over 2} \int _{t_0}^t
{\bf G}_h ^{(\lambda ) } (y, s, t)
\sum _{\mu \in \Lambda} | u_{\mu} (., s)|^2 ds$$
In order to solve such integral equations, we set, for each
interval $I= [t_0, t_1]$ ($0\leq t_0<t_1$):
$$E_1(I)=  \Big \{ u =
(u_{\lambda}) _{\lambda \in \Lambda} , \ \ \ \ u_{\lambda } \in
C^0 ( (\R^p)^{\Lambda} \times I, \hskip 1cm  \Vert u \Vert _{1,
I} = \sup _{ \lambda \in \Lambda \atop (x, t) \in (\R^p)^{\Lambda}
\times I} t^{-1} |u_{\lambda} (x, t)| < \infty  \Big  \}  , $$
and we denote by $B_{1, I}(r)$  the ball in $E_1(I)$ with radius $r$
and center at the origin.
First, we solve $u_y = T_y (u_y)$ in an interval $I = [0,
\tau]$, (the initial data vanishing at $t_0= 0$). We can choose $t_1>0$
 such that $T_y$ is a contraction from
$B_{1 , I}(2M_1(\e))$ into itself, where $I = [0, t_1]$ and
$M_1(\e)$ is defined in (1.7).
Let  $u _y= (u_{_y, \lambda })_{\lambda \in \Lambda}$
be the fixed point of  $T_y$ in this interval.
Then, we write the
integral equation that the second, and third order derivatives
must satisfy, and we solve them as the first one in an interval
$[0, t_1]$. It follows that the functions $u_{y , \lambda}$
are $C^2$ in $(\R^p)^{\Lambda } \times ]0, t_1 [$,
and satisfy the equation:
$${\partial (t u_{y, \lambda} ) \over
\partial t } \ +\ {x-y \over t} . \nabla (t u_{y, \lambda})  \ - \ {h^2 \over 2}
\Delta (t u_{y, \lambda} ) \ + \ \sum _{\mu \in \Lambda } ( u_{y, \mu }
(x, t)\ .\ \nabla _{x_{\mu  }} (t u_{y, \lambda} ) = t \partial
_{x_{\lambda}} V_{\Lambda} (x, t). \leqno (2.7) $$
The function $\psi _{\Lambda}$ defined by:
$$\psi _{\Lambda} (. , y, t) =
\int _0^t {\bf G}_h ( y, s, t)
\Big [ V_{\Lambda}   (.) -
{h^2 \over 2}  \sum _{\lambda \in \Lambda}
 | u_{y, \lambda} (. , s)| ^2  \Big ]  ds  $$
satisfies (2.1) in  $(\R^p)^{\Lambda } \times ]0, t_1 [$. The
maximum principle, (Proposition 2.1), applied to the equation
(2.7), shows that $|u_{y, \lambda } (x, t)| \leq {t \over 2}
M_1(\e)$ if $0\leq t \leq t_1$, which is a better estimation than
which is given by the fixed point theorem. Then, we take $\varphi
_{y, \lambda }(x) = u_{y, \lambda}(x, t_1)$ as an initial value
for a problem in an interval $I= [t_1 , t_2]$. We want a system of
functions, again denoted by $u_y$, such that $u_y= S_y \varphi_y +
T_y u_y$ in this interval. We can chose $t_2$ such that the map
$u\rightarrow S_y\varphi + T_y u $ is a contraction of  $B_{1
, I}(2M_1(\e))$ into itself, where $I = [t_1, t_2]$. We prove
again that  $u$ is $C^2$ on $(\R^p)^{\Lambda } \times ]t_1, t_2
[$, and satisfies (2.7), which implies again, by Proposition 2.1,
that $|u_{\lambda } (x, t)| \leq {t \over 2} M_1(\e)$ if $0\leq t
\leq t_2$. We can iterate this process, and prove the existence of
the functions $u_{y, \lambda}$ in a sequence of intervals $[t_j ,
t_{j+1}]$, in which these functions are $C^2$, with bounded
derivatives.  At each time, Proposition 2.1, applied to the
function $tu_{y , \lambda}$, to the equation (2.7), and to the
interval $[0, t_j]$ proves that the initial data $\varphi _{y,
\lambda}$ at time $t_j$ satisfies $\Vert  \varphi _{y, \lambda}
\Vert \leq {t_j \over 2} M_1 (\varepsilon) $. The length $t_{j+1}
- t_j$ of the interval,  allowing the fixed point theorem to work,
depends only of the  bound of this initial value, which is
independent of $j$. Therefore  $\tau = t_{j+1} - t_j$ may be
chosen independent of $j$. Thus, we proved the global existence of
$\psi _{\Lambda}$ of class $C^2$ satisfying (2.1), (2.2) and
(1.10). Then we solve the  integral equations satisfied by the
higher order derivatives, including the derivatives with respect
to $y$ also. For each order of derivation, we have to find a fixed
point for an integral operator, and the length of the intervals,
in which this operator may be contractive, depends only on the
bounds of the first order derivatives, already obtained globally.
Thus, we  prove the global existence of the higher order
derivatives, and that they are globally bounded in
$(\R^p)^{\Lambda} \times [0, T]$ for each $T>0$. This  provides
$\psi_\L$ verifying the properties stated in proposition 2.2.
\hfill \carre
\bigskip

Then the function
$U_{\Lambda }$ defined in (1.9), with this function $\psi _{\Lambda}$,
 is the integral kernel of the operator
$e^{-tH_{\Lambda }}$. Excepted for the first order derivatives,
the bounds given by the previous proof are not uniform with respect
to the set $\Lambda$.  The next step will be the proof of such uniform
bounds for the second order derivatives. For that,
we have to choose  a  norm $N_2$ on  the
 second differential of any function $f$ on $(\R^p)^{\Lambda}$. We
 set, for $\varepsilon \in ]0, 1[$:
 $$N_2 \Big ( d^2 f(x)  , \varepsilon \Big ) =
  \sup _{\lambda \in \Lambda } \sum _{\mu \in
\Lambda } {|\nabla _{x_{\lambda }}\nabla _{x_{\mu }}f(x) | \over
\varepsilon^{|\lambda - \mu |} }, \leqno (2.8)$$ (with the
euclidian norm on $\R^p$.) We denote by $\Vert \  \Vert $ the
$L^{\infty }$ norm.
 The norm $N_2$, and the  norms $N_m$ of (2.10),
 appear in  Sj\"ostrand [23] in the definition of
 {\it $0$-standard functions with exponential weight}. The norm
 $\Vert N_2 ( \psi_{\Lambda} (., y, t) \Vert $ is well defined
 for each $\Lambda$, $y$ and $t$, by Proposition 2.2. The norm
 $\Vert N_2 ( d^2V_{\Lambda}, \varepsilon) \Vert $ is bounded,
 independently of $\Lambda$, by the constant
 $M_2(\varepsilon)$ of (1.7).

 \bigskip \noindent {\bf Proposition 2.3.} {\it If $ht \leq
T_0 =  \Vert N_2 ( d^2 V_{\Lambda } (.) , \varepsilon) \Vert
^{-1/2}$, we
 have, for all $y \in (\R^p)^{\Lambda}$, with the notation (1.7),
 $$ \Vert N_2 ( d^2 \psi_{\Lambda } (.y, t)
  , \varepsilon) \Vert \leq 2 t  \Vert
  N_2 ( d^2 V_{\Lambda } (.) , \varepsilon)
  \Vert = 2t M_2(\varepsilon).  \leqno (2.9) $$ }

 \bigskip \noindent {\it Proof.} For each $\lambda \in \Lambda $,
for each $X$ in $\R^p$, for each sequence $(Y_{\mu })_{\mu \in
\Lambda }$ of vectors $Y_{\mu}$ in $\R^p$, and for each $y\in
(\R^p)^{\Lambda}$, the function $$ \varphi (x, t) = \sum _{\mu \in
\Lambda } {(X. \nabla _{x_{\lambda }}) (Y_{\mu }. \nabla _{x_{\mu
}}) \psi _{\Lambda }(., ., t, h ) \over \varepsilon ^{ |\lambda -
\mu |}}$$ satisfies the equation $L_a(t^2 \varphi )=  \ t^2 F -
h^2 t^2G$, where $L_a$ is defined in (2.3) with, here,
$a_{\lambda} = h^2 \nabla _{x_{\lambda}} \psi _{\Lambda }$ and:$$F
= \sum _{\mu \in \Lambda } {(X. \nabla _{x_{\lambda }})
 (Y_{\mu }. \nabla _{x_{\mu }})  V _{\Lambda }
\over \varepsilon ^{  |\lambda - \mu |}}   \hskip 1cm G(x) = \sum
_{(\mu , \nu ) \in \Lambda ^2 } {<\nabla _{x_{\nu}} (X. \nabla
_{x_{\lambda }})\psi _{\Lambda}(x, y, t)
 \ , \ \nabla _{x_{\nu}}  (Y_{\mu }. \nabla _{x_{\mu }})
\psi _{\Lambda }(x, y, t)>\over \varepsilon ^{ |\lambda - \mu
|}}$$ We have, by the definition (2.8)  of the norm $N_2$:
$$|F(x)| \leq N_2( d^2 V_{\Lambda}(x), \varepsilon)  \ |X| \ \sup
_{\mu \in \Lambda } |Y_{\mu }|, \hskip 1cm |G(x)| \leq N_2( d_x^2
\psi _{\Lambda} (x, y, t), \varepsilon)^2 |X| \ \sup _{\mu \in
\Lambda } |Y_{\mu }|.$$ Therefore, by the maximum principle,
(Proposition 2.1), applied to the function $u= t^2 \varphi$, and
to the operator $L_a$, with $t_0=0$, we have: $$\Vert t^2
\varphi(.,., t) \Vert \leq \ |X| \ \sup _{\mu \in \Lambda }
|Y_{\mu }| \  \int _0^t \Big [ s^2 \Vert N_2( d^2V_{\Lambda} (.) ,
\varepsilon ) \Vert + h^2s^2 \Vert N_2 (d_x^2 \psi _{\Lambda} (.,
y, s), \varepsilon)\Vert ^2  \Big ] \ ds .$$ In other words,
taking the sup on all vectors $X$ and sequences $Y_{\mu}$, and on
all points $x$ and $y$, $$t^2
 \Vert N_2 ( d_x^2 \psi _{\Lambda} (., y, t), \varepsilon )\Vert
 \leq \int _0^t
 \Big [ s^2
\Vert N_2( d^2V_{\Lambda} (.) , \varepsilon ) \Vert + h^2s^2 \Vert
N _2(d_x^2 \psi _{\Lambda} (., y, s), \varepsilon)\Vert ^2 \Big  ]
\ ds .$$
   Then Proposition 2.3 follows easily, by a kind of
quadratic Gronwall lemma.
\bigskip
\noindent {\it End of the proof of theorem 1.1.} Now, we need a
norm $N_m$ for the higher order differential $d^mf(x)$ of a
function on $(\R^p)^{\Lambda}$. We set: $$ N_m ( d^m f (x) ,
\varepsilon ) =\sup _{(\lambda _1, ..., \lambda_{m-1}) \in \Lambda
^{m-1}  } \sum _{\mu \in \Lambda} {\vert
  \nabla _{x_{\lambda _1}}...\nabla _{x_{\lambda _{m-1}}} \nabla _{x_{\mu
  }} f(x)\vert
\over
 \varepsilon ^{{\rm diam} ( \{ \lambda _1, ..., \lambda_{m-1}, \mu
 \})}}.
 \leqno (2.10)$$
 In [23], a family of functions, depending on $\Lambda$, is called
a $0-$standard function with exponential weight if $N_m (d^m f(x),
\varepsilon)$ is bounded, with bounds independent of $\Lambda$.
The definition of [23] is more complicated, since $\ell ^p$ norms
are used, not only $\ell ^1$ and $\ell ^{\infty}$ norms like here.
Again, $\Vert N_m ( d^m \psi _{\Lambda} (., ., t), \varepsilon )
\Vert$ is well defined by Proposition 2.2, and $\Vert N_m ( d^m V
_{\Lambda} , \varepsilon) \Vert $ is bounded, independently of
$\Lambda$,  if the hypothesis $(H_{\varepsilon})$ is satisfied.
Now, we can prove, by an induction on $m\geq 2$, that, if
($H_{\varepsilon }$) is satisfied, we have $$\Vert N_{m} ( d^m
\psi _{\Lambda} (., ., t), \varepsilon ) \Vert \leq t
K_m(\varepsilon ) \hskip 1cm {\rm if}\ \ \ \ ht \leq T_0, \leqno
(P_m)$$ where $K_m(\varepsilon)$ is independent of $\Lambda$. For
each sequence  $( \lambda ^{(1)}, ..., \lambda ^{(m-1)})$ of
points in $\Lambda$, for each vectors  $X_j$ ($1\leq j \leq m$)
and $Y_{\mu}$ $(\mu \in \Lambda)$ in $\R^p$, we shall estimate the
function $t^m \varphi$, where $$\varphi (x, t) = \sum _{\mu \in
\Lambda} {(X_1.\nabla_{x_{\lambda
_1}})...(X_{m-1}.\nabla_{x_{\lambda _{m-1}}})     (Y_{\mu
}.\nabla_{x_{\mu}}) \psi _{\Lambda} \over \varepsilon ^{ {\rm
diam} ( \{ \lambda_1, ..., \lambda_{m-1}, \mu \} )}}.$$ By
differentiating (2.1), we see that this his function satisfies an
equation of the type $L_a(t^m \varphi)=t^m F$ where $L_a$ is
defined in (2.3) with $a_{\mu} = h^2 \nabla _{x_{\mu}} \psi
_{\Lambda}$,  and  $F$ satisfies $$| F(x, y, t) | \leq\ \big [
\sup _{\mu \in \Lambda} | Y_{\mu}  |\big ] \ \left [  \prod _{ j=
1}^{m-1} | X_j | \right ]  \ \Bigg [ N_m( d^m V_{\Lambda} (x),
\varepsilon) +\ldots $$ $$ \ldots + C h^2 \sum _{k=2}^m N_k (
d_x^k (\psi _{\Lambda } (x, y, t) , \varepsilon) ) \
N_{m+2-k}d_x^{m+2-k} (\psi _{\Lambda } (x, y, t) , \varepsilon) )
\Bigg ].$$ We can apply the maximum principle (Proposition 2.1) to
the operator $L_a$
 as before. Taking the sup over all the sequences of vectors $X_j$ and
$Y_{\mu}$, using the hypothesis $(H_{\varepsilon})$ for
$V_{\Lambda}$, the induction hypothesis $(P_{m-1})$ for the terms
with $2<k<m$, and Proposition 2.3 for $k=2$ or $k=m$, we obtain,
if $ht \leq T_0$: $$ \Vert N_{m} ( d^m \psi _{\Lambda} (., ., t),
\varepsilon ) \Vert \leq (t + h^2 t^3) K_m(\varepsilon) + 4 C t
h^2 M_2(\varepsilon) \int _0^t N_{m} ( d^m \psi _{\Lambda} (., .,
s), \varepsilon )
 ds . $$ Then property $(P_m)$ follows by the usual Gronwall's
 lemma, (without more conditions on $h$ and $t$).

 \medskip Then, all derivatives with respect to $x$ only are
 bounded as claimed. Since $e^{-t H_{\Lambda}}$ is self-adjoint, we
 have also bounds for the derivatives with respect to $y$ only.
 For the mixed derivatives, we need a new induction. For example,
 to estimate the matrix $\nabla _{x_{\lambda}} \nabla _{y_{\mu}}
 \psi _{\Lambda } $, we see that, for $X$ and $Y$ in $\R^p$, the
 function $\varphi =<X .\nabla _{x_{\lambda}}> <Y . \nabla
 _{y_{\mu}}> \psi _{\Lambda }(x, y, t)$ satisfied $L_a(t \varphi ) = F$,
 where $L_a$ is defined in (2.3) and
 $$F =  <X .\nabla _{x_{\lambda}}> <Y . \nabla
 _{x_{\mu}}> \psi _{\Lambda } -h^2 \sum _{\nu \in \Lambda}
 \Big < \nabla _{x_{\nu}} <Y . \nabla _{y_{\mu}}> \psi _{\Lambda}\  .
 \ \nabla _{x_{\nu}} <X . \nabla _{x_{\lambda}}> \psi _{\Lambda}
 \Big >$$We use the estimations, already proven, of the
 derivatives with respect to $x$ only, we apply Proposition 2.1,
 and we apply again the usual Gronwall Lemma.
  The new induction follows  the same ideas, and
 leads to the proof of Theorem 1.1.

 \bigskip
\noindent {\bf 3.  Decomposition of a function on the
lattice.}
\bigskip

In this section, we shall define a decomposition of any function
$f$ on $(\R^p)^L$ (where $L= \Z^d$) as a sum of functions $T_Qf$
associated to the boxes of $L$. We shall give two variants of this
algorithm: the first one is simpler and works only for the
restriction of our integral kernel to the diagonal, and the second
one is useful for stronger estimates (for example, for trace
norms).  In the next section, we shall apply this decomposition to
the function $\psi _{\Lambda}$ of Theorem 1.1, and prove estimates
for the terms of the decomposition.
\bigskip
 We shall associate, to each function $f$ and to each box $Q$,
 a function $T_Qf$. If
$Q$ is a box in $L= \Z^d$, we denote by ${\rm Int}(Q)$ the set of
non empty boxes $Q'$ ($ Q' \subseteq Q$) which are either $Q$
itself, or obtained by removing in $Q$ some faces. We denote by
$m(Q, Q')$ the number of faces of $Q$ which are removed in $Q'$
($0 \leq m(Q , Q') < 2d$). If $Q$ is any set of $L= \Z^d$, we
define a map $\pi _Q : (\R^p)^L \rightarrow  (\R^p)^L $ by: $$(
\pi _Q x)_{\lambda} = \left  \{ \matrix { x_{\lambda } \hfill &
{\rm if} & \lambda \in Q \hfill \cr 0\hfill &{\rm if}& \lambda
\notin Q \hfill \cr } \right . \leqno (3.1)$$ Then, we define the
function $T_Qf$ by: $$(T_Qf) (x) = \sum _{ Q' \in {\rm Int}(Q)}
(-1)^{m(Q, Q')} \Big [ f( \pi _{Q'} x) -  f(0)\Big ] . \leqno
(3.2)$$
 Obviously, $T_Qf$ is supported in $Q$. If $f$ has a
finite support, let $\Lambda $ be a box containing the support of
$f$. Then we have: $$f(x) - f(0) = \sum _{ Q  \subseteq \Lambda }
(T_Qf) (x),  \leqno (3.3)$$ where the sum is taken over all the
non empty boxes $Q$ contained in  $ \Lambda$, including the
points. This equality follows from the definition of $T_Qf$ and
from the following remark, for each box $P$ contained in
$\Lambda$:$$\sum _{ Q \subseteq \Lambda \atop P \in {\rm Int} (Q)}
(-1)^{ m(Q , P)} = \left \{ \matrix { 1 \hfill & {\rm if}& P=
\Lambda \hfill \cr 0\hfill & {\rm if}& P \not= \Lambda \hfill \cr
} \right .$$

\medskip
In order to estimate $T_Qf$, we shall use,   rather than the
derivatives of $f$, some  operators of translation, with the
following notations.
 For each $u\in (\R^p)^L$, we set:
 $$(S _uf)(x) = f(x+u) -f(x)\hskip 1cm \sigma (u)= \{ \lambda \in \Z^d ,
 \ \ \ \ \ \
u_{\lambda } \not = 0 \} . \leqno (3.4) $$ For each box $Q$, of the form $Q=
\prod _{j= 1}^d [a_j , b_j]$, with $a_j \leq b_j$, we denote the
different "faces" of $Q$ by: $$B_-^{(k)}(Q)= \{ \lambda \in Q , \
\ \ \ \ \ \lambda _k= a _k\}, \hskip 1cm B_+^{(k)}(Q)= \{ \lambda
\in Q , \ \ \ \ \ \ \lambda _k= b _ k\} $$ For each
$k\leq d$ such that $a_k<b_k$, we can write $Q = [a_k, b_k] \times
\widetilde Q$, where $\widetilde Q$ is a box in $ \Z^{d-1}$. For
each box $P\in {\rm Int} \widetilde Q$, let us set: $$P_{\pm } =
[a_k, b_k] \times P, \hskip 1cm P_{+ } = ] a_k, b_k] \times P,
\hskip 1cm P_{- } = [a_k, b_k [ \times P, \hskip 1cm P_{0 } = ]
a_k, b_k[ \times P.$$
With these notations, we can write:
$$ T_Qf
(x) = \sum _{P \in {\rm Int} (\widetilde Q)} (-1)^{m( \widetilde Q
, P)} \Big [ f( \pi _{P _{\pm } } x ) -  f( \pi _{P _{ + } } x ) -
f( \pi _{P _{-} } x ) + f( \pi _{P _{0 } } x )  \Big ] .$$
We can
write this equality, using the previous operators $S_u$. If we set:
$ u_P =- \pi _{B_+^{(k)}(Q)}  \pi _{P _{\pm }} x$ and $ v_P =- \pi
_{B_-^{(k)}(Q)} \pi _{P_{\pm}} x$, we can write:
$$ f( \pi _{P
_{\pm } } x ) -  f( \pi _{P _{ + } } x ) - f( \pi _{P _{-} } x ) +
f( \pi _{P _{0 } } x )= ( S_{ u_P} S_{v_P} f ) ( \pi _{P_{\pm }}
x)  \leqno (3.5)$$
It follows that $$ \Vert T_Qf \Vert \leq 4^d \sup _{ \sigma
(u) \subset B_{+ }^{(k)} (Q) \atop \sigma (v) \subset B_{- }^{(k)}
(Q) } \Vert S _u S _v f \Vert. $$
   We have similar estimates for the derivatives   $\nabla
_{\lambda}T_Qf$, if $\lambda$ is neither in $B_{+ }^{(k)} (Q)$,
nor in $B_{- }^{(k)}(Q)$.  If $\lambda$ is, for example, in $B_{-
}^{(k)} (Q)$, we can use, instead of (3.5):$$ f( \pi _{P _{\pm } }
x ) - f( \pi _{P _{ + } } x ) - f( \pi _{P _{-} } x ) + f( \pi _{P
_{0 } } x ) = ( S_{ u_P} f) ( \pi _{P_{+ }} x) - ( S_{ u_P} f)(
\pi _{P_{\pm }} x),$$ and we get, if $\lambda$ is not in $B_{+
}^{(k)} (Q)$:$$ \Vert \nabla _{x_{\lambda}} T_Q f \Vert \leq 4^d
\sup _{ \sigma (u) \subset B_{+ }^{(k)} (Q) } \Vert S _u \nabla
_{x_{\lambda}} f \Vert. $$If $a_k= b_k$, we cannot apply that, but
we can write, in all the cases:$$ \Vert \nabla _{x_{\lambda}} T_Q
f \Vert \leq 4^d \Vert \nabla _{x_{\lambda}}  f \Vert.$$

\bigskip
It is possible to define similar projections by replacing in (3.2)
$\pi _Q$ by other projections. The heat  kernel and  our function
 $\psi_{\Lambda}$
  are defined  in $(\R^p)^{\Lambda} \times (\R^p)^{\Lambda}$, but we
  shall
 only use translations by vectors in the diagonal.
Therefore, we shall replace $\pi _Q$ by the following projector:
$$\Big ( \Pi _Q (x, y) \Big )_{\lambda} = \left  \{ \matrix {
(x_{\lambda }, y_{\lambda}) \hfill & {\rm if} & \lambda \in Q
\hfill \cr (0, y_{\lambda} - x_{\lambda}) \hfill &{\rm if}&
\lambda \notin Q \hfill \cr } \right . \leqno (3.6)$$ (We could
make another choice, preserving self-adjointness). For each
function $f(x, y)$ on $(\R^p)^L \times (\R^p)^L$, we define $T_Qf$
as before, with a small change: $$(T_Qf) (x, y) = \sum _{ Q' \in
{\rm Int}(Q)} (-1)^{m(Q, Q')} \Big [ f( \Pi _{Q'} (x, y))-  f(0,
y-x)\Big ].\leqno (3.7)$$ Then the operators $S _u$ are replaced
by the following one: $$(S_uf) (x, y) = f(x+u, y+u) - f(x, y)
\hskip 1cm \forall (x, y) \in (\R^p)^{L } \times (\R^p)^{L }\leqno
(3.8)$$ We verify easily the properties listed for later use in
the next proposition.

\bigskip
\noindent
{\bf Proposition 3.1.} {\it With the previous notations,
for each function $f$ in
$C^{\infty }( (\R^p)^{\Lambda } \times (\R^p)^{\Lambda })$, we have
\smallskip
\noindent i) $T_Qf$ depends only  on the variables $x_{\lambda }$
and $y_{\lambda }$ such that $\lambda \in Q$, and on the variables
$x_{\lambda} -y_{\lambda }$ such that $\lambda \notin Q$. If $f$
is supported in a finite box $\Lambda$, we have: $$f(x, y)- f(0,
y-x) = \sum _{Q \subseteq \Lambda } (T_Qf)(x, y) \leqno (3.9)$$
ii) If $f$ depends in a smooth way on a parameter $\theta$, we
have $T_Q ({\partial f \over \partial \theta})= {\partial (T_Qf)
\over
\partial \theta}$.
\smallskip
\noindent iii) If $\lambda ^{(1)}$, ... $\lambda ^{(m)}$ is a
finite sequence of points in $\Lambda $, we have: $$\Vert \nabla
_{\lambda ^{(1)}} ... \nabla _{\lambda ^{(m)}} T_Qf \Vert  \leq
4^d \Vert \nabla _{\lambda ^{(1)}} ... \nabla _{\lambda ^{(m)}} f
\Vert . \leqno (3.10)$$
iv) If moreover, none of the points
$\lambda ^{(j)}$ is in the face $B_{+ } ^{(k)} (Q)$, we have $$\Vert \nabla
_{\lambda ^{(1)}} ... \nabla _{\lambda ^{(m)}} T_Qf \Vert  \leq
4^d \sup _{ \sigma (u) \subset B_{+ }^{(k)} (Q)}\Vert \nabla
_{\lambda ^{(1)}} ... \nabla _{\lambda ^{(m)}} S_u f \Vert .
\leqno (3.11)$$
v) If $a_k \not= b_k$, and if  the points $\lambda
^{(j)}$ are neither in $B_{+ } ^{(k)} (Q)$, nor in $B_{- } ^{(k)}
(Q)$ we have $$\Vert \nabla _{\lambda ^{(1)}} ... \nabla _{\lambda
^{(m)}} T_Qf \Vert \leq 4^d \sup _{ \sigma (u) \subset B_{+
}^{(k)} (Q) \atop \sigma (v) \subset B_{- }^{(k)} (Q) }\Vert
\nabla _{\lambda ^{(1)}} ... \nabla _{\lambda ^{(m)}} S_u S_vf
\Vert . \leqno (3.12)$$ }

\bigskip
\noindent {\bf 4. Proof of Theorem 1.2, excepted (1.15).}
\bigskip
The functions $T_Q \psi _{\Lambda}$ are defined by (3.7), and the
properties 1 and 2 of Theorem 1.2 follow from Proposition 3.1. It
remains to prove the bounds. We shall modify the norm $N_m (d^m
f(x), \varepsilon)$ of section 2, and now, it will depend also on
one subset $E$ of $\Z^d$, or sometimes on two subsets $E$ and $F$.
We set,
for each $E_1$, $\ldots$ , $E_m$ which are either subsets of
$\Z^d$, or points of $\Z^d$, $$D( E_1 , \ldots , E_m) =
 \sup _{j , k \leq m } {\rm dist} (E_j , E_k). \leqno (4.1)$$
  If $f$ is a $\C$ function on
   $(\R^p)^{\Lambda} \times (\R^p)^{\Lambda}$, we set, for
   each $m\geq 2$,  for each point $(x, y)$, for each
   $\varepsilon $ in $]0, 1[$, and for each subsets $E$ and $F$
    of $\Z^d$ :
   $$N_m (d^m f (x, y), \varepsilon , E, F)
    =\sup _{(\lambda ^{(1)}, \ldots ,
\lambda ^{(m-1)})\in \Lambda^{m-1}  } \sum _{\mu \in \Lambda } {
\vert \nabla _{ \lambda ^{(1)}} \ldots \nabla _{ \lambda ^{(m-1)}}
\nabla _{ \mu} f(x, y) \vert \over \varepsilon ^{  D(\lambda
^{(1)}, \ldots , \lambda ^{(m-1)}, \mu , E, F )}}.\leqno (4.2)$$
If $m=1$, there is no sup in (4.2), and the sum is taken on
all $\lambda \in \Lambda$. We set also, if $m\geq 1$:
$$N_m ^{\infty } (d^m f
(x, y), \varepsilon , E, F)
    =\sup _{(\lambda ^{(1)}, \ldots ,
\lambda ^{(m)})\in \Lambda^{m}  } { \vert \nabla _{ \lambda
^{(1)}} \ldots \nabla _{ \lambda ^{(m)}}  f(x, y) \vert \over
\varepsilon ^{  D(\lambda ^{(1)}, \ldots , \lambda ^{(m)}, E, F
)}}.\leqno (4.3)$$  When no set $E$ or $F$ appear, the first  norm
$N_m$  is the same as in section 2 if $m\geq 2$. For $m=1$,
the sup in section 2 (without sets $E$ and $F$)
is repaced by a sum here (when there is at least one set).
The second norm
$N_m^{\infty}$ will be used in section 6.  We still denote by
$\Vert \ \Vert$ the $L^{\infty}$ norm. With these notations, the
estimation (1.14) of Theorem 1.2 will follow from the next Lemma.

\bigskip
\noindent {\bf Lemma 4.1.} {\it If $(H_{\varepsilon})$ is
satisfied $(0 < \varepsilon < 1)$, for each
integer $m\geq 1$, there exists $K_m(\varepsilon )>0$ such that,
for each finite box $\Lambda $ of $\Z^d$, for each vectors $u$ and
$v$ in $(\R^p)^{\Lambda}$,  the functions $ S_u \psi _{\Lambda}$
and $ S_u S_v \psi _{\Lambda}$ defined like in (3.8), (with the
function $\psi _{\Lambda}$  of Theorem 1.1), satisfy, if $ht \leq
T_0$, (defined in (1.7) and (1.8)):
$$\Vert N_m( d^m \big (S_u \psi _{\Lambda}\big ) (., . , t) ,
\varepsilon, \sigma (u) )\Vert \leq t K_m (\varepsilon ) \vert
\sigma (u) \vert , \leqno (4.4)$$
$$\Vert N_m( d^m \big (S_u S_v
\psi _{\Lambda}\big ) (., . , t) , \varepsilon, \sigma (u), \sigma
(v) )\Vert  \leq t K_m (\varepsilon )
  | \sigma (u)|  | \sigma (v) |, \leqno
(4.5)$$
where the support $\sigma (u)$ is defined in (3.4).
If $\sigma (u) \cap \sigma (v) = \emptyset $, we can write
also:
$$| S_u S_v \psi _{\Lambda }(x , y, t)| \leq t
K_0(\varepsilon) \varepsilon ^{{\rm dist} (\sigma (u) , \sigma
(v))}  | \sigma (u)| | \sigma (v) |. \leqno (4.6) $$
 }

 \bigskip
    Since one argument of the proof will be used again twice in
 section 6, we state it as a Lemma.

 \bigskip
\noindent {\bf Lemma 4.2.} { \it We consider a real valued
function $A(x, y, t)$, $\C$ on $(\R^p)^{\Lambda} \times
(\R^p)^{\Lambda} \times [0, \infty [$. We assume that there is
$\varepsilon $ in $]0, 1[$  such that, for each $m\geq 1$, we can
write, for some constant $K_m>0$:
$$ \Vert N_m (d^m A(., t) ,
\varepsilon)\Vert  \leq t K_m, \leqno (4.7) $$
where $N_m$ is the
norm of section 2. We denote  $L_a$ is the differential operator
of (2.3), with  $a_{\lambda }=h^2 \nabla _{x_{\lambda}} A$. We
consider also a smooth, real valued function $\varphi (x, y, t)$,
such that $\varphi (y, x, t) = \varphi (x, y, t)$ and $\varphi (x,
y, 0)=0$. We assume that there are two subsets $E$ and $F$ of
$\Z^d$ such that, for each $m\geq 1$,

$$ \Vert N_m (d^m ( L_a\varphi) (. , ., t)  ,
\varepsilon, E, F) \Vert \leq K_m.  \leqno (4.8)$$
Then, if $ht$ is bounded,  we can write, with
another $K_m$, for each $m\geq 1$:$$\Vert N_{m} (d^m \varphi (.,
., t), \varepsilon, E, F)\Vert  \leq
 t K_m . \leqno (4.9)$$
 We have the same result with only one set $E$. We have
 also the same result if we replace, for all $m$,  the norm $N_m$ by the norm
 $N_m^{(\infty)}$, both in the hypothesis (4.8) and in the conclusion
 (4.9).
}
\bigskip
The proof is exactly like in section 2, but simpler since we have
no more quadratic Gronwall Lemma, but the usual one. (In section
2, we had $A = \psi _{\Lambda}$ and $\varphi  $ was a derivative of
$A$.  We had to prove, in the same time, bounds for $A$ and for
$\varphi $, but now the bounds for $A$ are already available.)

\bigskip
\noindent {\it Proof of Lemma 4.1.} Since $\psi _{\Lambda}$
satisfies (2.1), if we apply to it the operator $S_u$ ( $u$ in
$(\R^p)^{\Lambda}$) defined in (3.8), the function $f  =
S_u\psi _{\Lambda }$
 satisfies, as a function of $x$
and $t$, while $y$ and $u$  are fixed, an equation of the form
$L_a (f ) = F$, where $L_a$  is the differential operator
defined in (2.3), with: $$a_{\lambda} (x, y, t) ={h^2\over 2}
\nabla _{x_{\lambda}} \left [ \psi _{\Lambda }(x+ u, y+u, t) +
\psi _{\Lambda } (x, y, t)\right ] \hskip 1cm F(x) = S_u
V_{\Lambda } (x). \leqno (4.10) $$
By the hypothesis $(H_{\varepsilon})$, we can
write, for each $m\geq 1$, with some constant $K_m(\varepsilon)$
independent of $u$ and $\Lambda$: $$\Vert N_m (d^m (S_u
V_{\Lambda} )(., ., t) , \varepsilon, \sigma (u))\Vert \leq  K_m
(\varepsilon ) \vert \sigma (u) \vert .$$ Then, (4.4) follows from
Lemma 4.2. For the proof of (4.5), we remark that, if
$f = S_u \psi _{\Lambda}$ satisfies $L_a f = F$, where $a$ and $F$
are defined in (4.10),  the new function
 $g = S_u S_v \psi_{\Lambda }= S_vf$ satisfies
 $L_{b} g  = G$, with:
 $$b_{\lambda} (x,y, t) = a_{\lambda} (x+v , y+ v, t)
 \hskip 1cm G = S_u S_v V _{\Lambda} - \sum _{\lambda \in \Lambda}
 (S_v a_{\lambda} ). ( \nabla _{x _{\lambda }}  S_u \psi
 _{\Lambda}) .$$
  By the hypothesis $(H_{\varepsilon})$, and by (4.4),
for each $m\geq 1$, we can write, if$ht \leq T_0$:
$$\Vert N_m (d^m (G)(., ., t) , \varepsilon, \sigma (u), \sigma (v) )\Vert
\leq  K_m(\varepsilon) \  \vert \sigma (u)\vert \  \vert \sigma (v)
\vert . $$
Then (4.5) follows from Lemma 4.2.
If $m=0$, we remark that, if $\sigma (u) \cap \sigma (v) = \emptyset $,
the unbounded self-interaction term disappear in $S_u S_v \psi
_{\Lambda}$, and therefore $\Vert
S_u S_v V _{\Lambda }\Vert \leq K(\varepsilon) \varepsilon ^{ {\rm
dist} ( \sigma (u), \sigma (v) ) } \inf ( \vert \sigma (u)\vert,
\vert \sigma (v) \vert )$. Then (4.6) follows from Proposition
2.1.
\medskip
\noindent {\it Proof of the estimations of Theorem 1.2.}  If $m=0$
and ${\rm diam}(Q) \not= 0$, the bound (1.) follows from  the
point v) of Proposition 3.1 and (4.6), remarking that $\vert
B_{\pm }^{(j)} (Q)\vert \leq (1 + {\rm diam}(Q) )^{d-1}$. If
$m\geq 1$, according to the geometric situation of the points
$\lambda ^{(j)}$ and the box $Q$, we  apply either the point v) of
Proposition 3.1 and (4.5), or the point  iv) and (4.4), or the
point iii) and Theorem 1.1.
\bigskip
When $Q$ is reduced to a single point $\lambda $,  $T_Q \psi
_{\Lambda }$ is not bounded, and we got  bounds only for its
derivatives. Some information on $T_{\{ \lambda \} } \psi
_{\Lambda}$ itself will be given in the next section.

\bigskip
\noindent {\bf 5. Splitting the set $\Lambda$: case of a small
subset. }
\bigskip
In applications to the decay of correlations, (section 8),  the
set $\Lambda$ will be a large box,  containing a smaller set $E$,
(the union of the two supports of the observables), and we have to
dissociate the two sets $E$ and $\Lambda \setminus E$.  If $Z
_{\Lambda}(t)$ is the partition function of (1.2), the next
proposition will be useful, for example, to compare $Z
_{\Lambda}(t)$ and $Z _{\Lambda \setminus E}(t)$, and to estimate the
trace norm of the partial trace (with respect to $E$), of some
operators.
\bigskip
For each operator $K$ with an integral kernel $K(x, y)$ in
the Schwartz space ${\cal S} ( (\R^{2p})^{\Lambda} \times
(\R^{2p})^{\Lambda} )$, for each subset $E \subset \Lambda$,
for each $m$, $m'$ and $\mu$, we set:
$$ \Vert K \Vert _{ m , m' , \mu} =
 \sup _{\lambda ^{(1)},
... ,\lambda ^{(k)} \in E^k \atop k \leq m}\
 I _{  m' , \mu} (
  \nabla _{\lambda ^{(1)}} ... \nabla _{\lambda ^{(k)}} K),
  \leqno (5.1)$$
where $$I _{  m' , \mu}  (K) =
  \int (1 +|x_E|_1 + |y_E|_1)^{m'} \ e^{\mu |x_E - y_E|_1} \  \Big  |
 K( x_E , x_{\Lambda \setminus E} , y_E ,
x_{\Lambda \setminus E})  \Big | dx_E dy _E dx_{\Lambda \setminus
E}.$$

\bigskip
\noindent {\bf Proposition 5.1.}  { \it With these notations, we
can write, for each $m$, $m'$ and $\mu $, for some constant $M(
|E| , t, h, \varepsilon )$, $$ \Vert  U _{\Lambda} (t) \Vert _{ m
, m', \mu } \leq M( |E| , t, h, \varepsilon ) \ Z_{\Lambda }
(t).$$ }
\bigskip
The proof relies on the next Proposition, also used for the point
(1.15) of Theorem 1.2.
  If $E \subset \Lambda \subset \Z^d$, we denote by $\psi _{\Lambda
\setminus E} $ the function defined by Theorem 1.1, for the set
$\Lambda \setminus E$. It can be seen also as a function on
$(\R^p)^{\Lambda} \times (\R^p)^{\Lambda} \times \R_+$, depending
only on $x_{\Lambda \setminus E}$, $y_{\Lambda \setminus E}$, and
$t$. For each point $\lambda \in E$, we set $A_{\lambda }(x)=
A(x_{\lambda})$, where $A$ is the self-interaction term in the
definition (1.4), and we denote by $\widetilde A_{\lambda}(x, y)$
the mean value of $ A _{\lambda}$ in the segment between $x$ and
$y$, as in (1.16).

\bigskip
\noindent {\bf Proposition 5.2.} {\it  With these notations, there
exists $K(\varepsilon)>0$, independent of the two sets $E$ and
$\Lambda $ ($E \subset \Lambda \subset \Z^d$), such that: $$  \Big
\Vert \psi _{\Lambda } (. , ., t) -\Big  [ t \sum _{\lambda \in E}
\widetilde A _{\lambda} + \psi _{\Lambda \setminus E} (. ,
.,t)\Big ]  \Big \Vert \leq   \vert E \vert K(\varepsilon) (t +
h^2 t^2). \leqno (5.2)$$ }
\bigskip
\noindent {\it Proof.  First step.}
 Let us denote by $V_{\rm disc }$ the
potential defined as if all the points of $E$ had no interaction
with the others points in $\Lambda $, and no interactions between
themselves: $$V_{\rm disc } = \sum _{\lambda \in \Lambda }
A_{\lambda } \ +\ \sum _{\lambda , \mu \in \Lambda \setminus E}
B_{\lambda , \mu }.$$ For each $\theta \in [0, 1]$, let us set
$V_{\Lambda , E , \theta} = V_{\rm disc } + \theta (V_{\Lambda } -
V_{\rm disc } )$. Let us denote by   $\psi _{\Lambda , E , \theta}
$ the solution of the Cauchy problem (2.1), (2.2), where
$V_{\Lambda}$ is replaced by  $V_{\Lambda , E , \theta}$. We see
easily that:
 $$\psi _{\Lambda , E, 0} = \sum _{\lambda \in
E} \psi _{ \{ \lambda \} } (x_{\lambda }, y_{\lambda } , t ) +
\psi _{ \Lambda \setminus E} (x_{\Lambda \setminus E} , y_{\Lambda
\setminus E}, t)\ \ \ \ \ \ \psi _{\Lambda , E, 1} =\psi
_{\Lambda} , \leqno (5.3) $$where $\psi _{ \{ \lambda \} }$ is the
function defined by Theorem 1.1 for the set reduced to the point $
\{ \lambda \}$, and to the potential $V_{ \{ \lambda \} } =
A_{\lambda}$ reduced to the self-interaction term, according to
(1.4). If we differentiate with respect to $\theta$ the non linear
equation like (2.1), satisfied by $\psi _{\Lambda , E , \theta}$
(with $V_{\Lambda , E , \theta}$ in the RHS) , we see that $L_a
{\partial \psi _{\Lambda , E , \theta} \over
\partial \theta} =F$, where $L_a$ is the differential operator
of (2.3),  with $a_{\mu}= h^2 \nabla _{x_{\mu}} \psi _{\Lambda , E
, \theta}$ and with  $F = V_{\rm disc}-V_{\Lambda}$. By the
hypothesis $(H_{\varepsilon})$, we can write : $\Vert V_{\Lambda }
- V_{\rm disc} \Vert \leq K(\varepsilon ) \vert E\vert $.
 Therefore, by Proposition 2.1, we can write $\Vert  {\partial
\psi _{\Lambda , E , \theta} \over
\partial \theta} (., ., t)\Vert \leq t K(\varepsilon) \vert
E\vert$. Therefore we have,  with another $ K(\varepsilon )$,
$$\Vert \psi _{\Lambda } - \psi _{\Lambda , E, 0 } \Vert \leq  t \
\vert E\vert \ K(\varepsilon). \leqno (5.4) $$

\medskip
\noindent {\it Second step.} Now, we shall compare $\psi _{ \{
\lambda \} } $, associated by Theorem 1.1 to the single set $\{
\lambda \}$,  and its semi-classical approximation $\psi _{ \{
\lambda \} }^{(0)} (x, y, t) =  t \widetilde A(x_{\lambda},
y_{\lambda})$. By a direct computation, we see that $${\partial
\psi_{ \{\lambda \} }^{(0)} \over
\partial t} + {x-y\over t} \cdot \nabla _x  \psi_{ \{ \lambda \}
}^{(0)} = A _{\lambda} $$  and therefore that the function
$\varphi =\psi_{ \{ \lambda \} }-\psi_{ \{ \lambda \} }^{(0)}$
satisfies the equation $L_0( \varphi ) = F$ where $L_0$ is the
operator defined in (2.3,  with $a_{\mu } = 0$ and $F = {h^2 \over
2}( \Delta _x \psi_{ \{ \lambda \} }^{(0)} -  \vert \nabla _x
\psi_{ \{ \lambda \} }\vert ^2)$. By Theorem 1.1, we can write, if
$ht \leq T_0$, $\Vert F(., ., t)\Vert \leq K(\varepsilon )  h^2 (t
+ t^2)$. By Proposition 2.1, it follows that: $$ \Vert \psi _{ \{
\lambda \} } -     \psi _{ \{ \lambda \} }^{(0)} (., ., t) \Vert
\leq K(\varepsilon ) h^2 (t^2 + t^3) \hskip 1cm {\rm if}\ \ \ \ ht
\leq T_0 \leqno (5.5)$$where $T_0$ is the constant of (1.8) and
(1.7). Proposition 5.2 follows from (5.3), (5.4)  and  (5.5).
 \bigskip
\noindent {\it Proof of the point (1.15) of Theorem 1.2.} We apply
Proposition 5.2 with  $E$  reduced to a single point $\lambda$.
Then, we apply (3.10) with $m=0$, $Q = \{ \lambda \}$, and $f$
being the function in the LHS of (5.2). Thus we obtain: $$\Vert
T_{ \{ \lambda \} } \psi _{\Lambda } (. , ., t) - t T_{ \{ \lambda
\} } \widetilde A_{\lambda}  - T_{ \{ \lambda \} } \psi
_{\Lambda\setminus \{ \lambda \}  } (. , ., t) \Vert \leq
K(\varepsilon) (t + h^2 t^2 ).$$ Using the definition (3.7) of
$T_{ \{ \lambda \} }$, we see that $$T_{ \{ \lambda \} } \psi
_{\Lambda \setminus \{ \lambda \} }=0 \hskip 1cm T_{ \{ \lambda \}
} \widetilde A_{\lambda} (x , y)
 =\widetilde A (x_{\lambda} , y_{\lambda}) -
  \widetilde A(0, y_{\lambda } - x_{\lambda}),$$
 and the bound (1.15) of Theorem 1.2 follows.

\bigskip We may also apply Proposition 5.2 to $E =
\Lambda$, and we obtain that,  if $ht \leq T_0$, the integral
kernel $U_{\Lambda} (x, y, t)$ is in ${\cal S}( (\R^p)^{\Lambda}
\times (\R^p)^{\Lambda} )$, but of course, in this way, all the
bounds will depend on $\Lambda$.

  \bigskip
\noindent {\bf  6. Splitting the box $\Lambda$ by an hyperplane.}
\bigskip
This section will be used in the proof of Theorems 1.4 and 1.5.
 Let  $\Lambda \subset \Z ^d$ be the union of two disjoints subsets
$\Lambda _1 $ and $\Lambda _2$, which  are separated by an
hyperplane $\Sigma$, orthogonal to one of the vectors of the
canonical basis. We may assume that, for some $j\leq d$ and
$\sigma \in \Z$: $$ \Sigma = \{ \lambda \in \Z^d, \ \ \ \ \lambda
_j = \sigma \}, \ \ \ \ \ \ \Lambda _1 \subset \{ \lambda \in
\Z^d, \ \ \ \ \lambda _j \leq \sigma \},
 \ \ \ \ \ \ \Lambda _2
\subset  \{ \lambda \in \Z^d, \ \ \ \ \lambda _j \geq  \sigma  \}.
\leqno (6.1)$$
We have again, like in Section 5, to dissociate the two subsets,
but now, we need, in the situation (6.1), estimations which are
uniform with respect to both of them.
 We denote by
$V_{\rm inter}$ the sum of the interactions between a point of
$\Lambda _1$ and a point of $\Lambda_2$ in the definition (1.4) of
$V_{\Lambda}$:$$V_{\rm Inter}(x)= \sum _{\lambda \in \Lambda _1,
\mu \in \Lambda _2} B_{\lambda , \mu }(x_{\lambda }, x_{\mu}).
\leqno (6.2) $$For each $\theta \in [0, 1]$, we set $V_{\Lambda ,
\theta } =  V_{\Lambda } - \theta V_{\rm Inter }$, and we denote
by $H_{\Lambda , \theta }$ the Hamiltonian defined as in (1.1),
with $V_{\Lambda }$ replaced by $V_{\Lambda , \theta }$, and by
$\psi_{\Lambda , \theta}$ the function associated to this Hamiltonian
by Theorem 1.1.
We are now interested to the derivative of
this function with respect to $\theta$, and to the decomposition
$T_Q{\partial \psi _{\Lambda ,\theta } \over\partial \theta }$
defined in section 3. For each box $Q$ of $\Z^d$, let $\pi _{\Sigma}(Q)$
be the orthogonal projection of $Q$ on $\Sigma$.

\bigskip
\noindent {\bf Proposition 6.1.} {\it With the previous notations,
we can write, for some constant $C_m(\varepsilon)$ independent of
$\Lambda$, $\Lambda _1$ and $\Lambda _2$,
   for each points $\lambda ^{(1)}$, ...
$\lambda  ^{(m)}$ in $\Lambda $, for each box $Q \subseteq \Lambda
$,  if $ h t \leq T_0$    and $\theta \in [0,
1]$:
$$\Vert \nabla _{\lambda ^{(1)}} ... \nabla _{\lambda ^{(m)}}
\Big (T_Q {\partial \psi _{\Lambda , \theta} \over
\partial \theta }(. , ., t)   \Big )  \Vert \leq tC_m (\varepsilon)
\varepsilon ^{ {\rm diam}  (Q\cup \pi _{\Sigma}(Q)  \cup \{
\lambda ^{(1)} , ..., \lambda^{(m)} \})} \ \big (1 + {\rm diam
}(Q) \big )^d \leqno (6.3) $$  }
 \bigskip
 Since $V_{\Lambda , \theta }$ is of the same type that $V_{\Lambda }$,
and satisfies the same hypotheses, with bounds independent of
$\theta$, (excepted the hypothesis (1.6), which is not needed
for  theorems 1.1 and 1.2), the bounds given by Theorem 1.1, 1.2, and Lemma 4.1 for
the function $\psi _{\Lambda , \theta}$ are also uniform in
$\theta$. With this remark, the proof of Proposition 6.1 relies
on the following Lemma.

 \bigskip
\noindent {\bf Lemma 6.2.} {\it For each integer $m\geq 1$, there
exists $K_m( \varepsilon)>0$  such that,  if $ ht \leq T_0$, we
have,  for each points $\lambda ^{(1)}$, $\ldots $ $\lambda
^{(m)}$ in $\Lambda $:

$$ \Vert \nabla _{\lambda ^{(1)}}\ldots \nabla _{\lambda ^{(m)}}
{\partial \psi _{\Lambda , \theta }\over
\partial \theta } \Vert \leq
K_m (\varepsilon) \ t \ \varepsilon ^{ D(\lambda ^{(1)},... ,
\lambda ^{(m)}, \Sigma ) }.  \leqno(6.4) $$
 Moreover,
for each $u$  in $(\R ^p)^{\Lambda }$ such that the support  $\sigma (u)$
 is contained, either in $\Lambda _1$, or in $\Lambda _2$,
we have the following bound, valid also for $m=0$ :
$$ \Vert
\nabla _{\lambda ^{(1)}}\ldots \nabla _{\lambda ^{(m)}} S_u
{\partial \psi _{\Lambda , \theta }\over
\partial \theta } \Vert \leq K_m (\varepsilon) \ t \ \varepsilon ^{
D(\lambda ^{(1)},... , \lambda ^{(m)}, \sigma (u), \Sigma ) }\
 | \sigma (u) | , \leqno(6.5) $$
 where
 $D(\lambda^{(1)}, ..., \lambda^{(m)}, \Sigma)$ is defined in (4.1).
 }

\bigskip
\noindent {\it Proof.} Since $\psi _{\Lambda , \theta}$ satisfies
(2.1), with $V_{\Lambda}$ replaced by $V_{\Lambda , \theta}$, it
follows that $f = {\partial \psi _{\Lambda ,\theta } \over\partial \theta }$
satisfies $L_a f  = - V_{\rm Inter}$, where $L_a$ is the
operator of (2.3), and $a_{\lambda} =h^2 \nabla _{x_{\lambda}} \psi
_{\Lambda ,\theta}$. We can write, for each $m\geq 1$:
$$ \Vert
N_m^{\infty} ( d^m V_{\rm Inter} , \varepsilon, \Sigma ) \Vert
\leq K_m(\varepsilon ).$$
Then (6.4) follows from Lemma 4.2, with
the $ N_m^{\infty} $ norm.
Now, we apply the operator $S_u$ defined  in (3.8)
to $ f= \partial _{\theta} \psi_{\Lambda , \theta}$.
Since $L_a f  = - V_{\rm Inter}$, it follows
 that the function
$g = S_u f$ satisfies $L_a g = G$, where still
$a_{\lambda} =h^2 \nabla _{x_{\lambda}} \psi _{\Lambda ,\theta}$
and:
$$ G(x, y, t)  = - S_u V_{\rm Inter}(x) - h^2\sum _{\lambda \in
\Lambda}  \Big ( \nabla _{x_{\lambda}}  S_u \psi
_{\Lambda ,\theta } \Big )\ . \ \Big ( \nabla _{x_{\lambda}}
\partial _{\theta} \psi _{\Lambda , \theta}
 (x+u, y+u, t) \Big ) .$$
In order to estimate this function, we use
the form (6.2) of $V_{\rm Inter}$, the hypothesis
$(H_{\varepsilon})$, the bound (6.4), and  Lemma 4
(bound (4.4)), applied to $\psi _{\Lambda, \theta}$
instead of $\psi _{\Lambda}$. We obtain, if $ht \leq T_0$:
$$\Vert G (., ., t)  \Vert \leq
K_0(\varepsilon) \varepsilon ^{ {\rm dist} ( \sigma (u), \Sigma)} |\sigma
(u)| .$$
By the same arguments, we can write, for each $m\geq 1$,
$$\Vert N_m^{\infty}(d^m G (., .,
t) , \varepsilon , \sigma (u), \Sigma)\Vert   \leq
K_m(\varepsilon) |\sigma (u)| .$$
Then, for $m=0$, the estimation (6.5) follows from Proposition 2.1
and, for $m\geq 1$,
it follows from Lemma 4.2 (with the $N_m^{\infty}$ norm).

\bigskip
\noindent {\it End of the proof of Proposition 6.1.} We may assume
that $\Lambda _1$, $\Lambda _2$ and $\Sigma$ satisfy (6.1), with
$j=1$. If the box $Q \subset \Lambda _1 \cup \Lambda _2$ is not
disjoint from $\Sigma$, (6.3) follows from Theorem 1.2. Now, we
may assume that $Q = \prod _{k= 1}^d [a_k, b_k]$, with $\sigma  <
a_1 \leq b_1$ and $a_k \leq b_k$ for $k \geq 2$. Let $k\leq d$
such that the diameter of the set $Q\cup \pi _{\Sigma}(Q)  \cup \{
\lambda ^{(1)} , ..., \lambda^{(m)} \}$ is the length of its
$k-$th projection. If $k \not= 1$, the estimation (6.3) follows
from Theorem 1.2 (with the potential $V_{\rm Inter}$). If $k=1$,
and $\lambda _1^{(k)} \not= b_1$ for all $k$, it follows from
(6.5) and  the point iv of Proposition 3.1, applied with the set
$B_1^+ (Q)$. In the other case, it follows from (6.4) and point
iii) of Proposition 3.1.

\bigskip

\bigskip
\noindent {\bf 7. Representation of the quantum correlations.  }
\bigskip
In this section, we shall give an expression of the correlation
${\rm Cov}  _{\Lambda , t}(A,B)$, (see (7.4)- (7.5)). Then, we
shall prove Theorem 1.3, point b, i.e.  when the local observables
$A$ and $B$ are multiplications. In section 8, we shall prove
Theorem 1.3 in the general case. For the expression of the
correlation, we shall use the tensor product of the heat kernel by
itself (doubling of variables), and the action, on this kernel, of
a finite group of symmetries.

 \bigskip
 \noindent {\it 1. Doubling of variables. } We shall denote  by
$(x', x'')$ the variable of $(\R^p)^{\Lambda } \times
(\R^p)^{\Lambda }$.  For each operator $A$ in $L^2((\R^p)^{\Lambda
})$, we denote by $A'$ (resp. $A''$ ) the operator $A$, seen as an
operator in $L^2((\R^p)^{\Lambda } \times (\R^p)^{\Lambda })$,
acting only on the variable $x'$ (resp. on $x''$).  We denote by
$\widetilde H_{\Lambda}$ the operator in $(\R^p)^{\Lambda } \times
(\R^p)^{\Lambda }$ defined by $\widetilde {H_{\Lambda }}=
H'_{\Lambda } + H''_{\Lambda} $. We denote by ${\rm
Tr}_{\Lambda}(A)$ the trace of an operator $A \in {\cal L}
(\widetilde {\cal H}_{\Lambda})$, where  $ \widetilde {\cal
H}_{\Lambda} = L^2( (\R^p)^{\Lambda} \times  (\R^p)^{\Lambda}$.
Then, an easy computation shows that, if $A\in {\cal L} ({\cal H}
_{E_1})$ and $B\in {\cal L} ({\cal H} _{E_2})$, where $E_1$ and
$E_2$ are disjoint subsets, contained in a same box $\Lambda$, we
have: $${\rm Cov} _{\Lambda , t } (A, B) = { 1\over 2 \widetilde
Z_{\Lambda}(t) } \ {\rm Tr}_{\Lambda} \ \left ( e^{-t \widetilde
H_{\Lambda } } (A' -A'') (B' - B'') \right ) \hskip 1cm \widetilde
Z_{\Lambda} (t) \ = \ {\rm Tr}_{\Lambda} \ \left ( e^{-t
\widetilde H_{\Lambda } } \right ). \leqno (7.1)$$

\bigskip

\noindent {\it 2. Group of symmetries, and averaging.} If
$\psi_{\Lambda }$ is the function, associated by Theorem 1.1, to
the operator $H_{\Lambda}$, and  $\widetilde \psi _{\Lambda }$ to
the operator $\widetilde {H_{\Lambda}}$, we  have:$$\widetilde
\psi _{\Lambda }(X, t) = \psi_{\Lambda}(x' , y', t) +
\psi_{\Lambda}(x'' , y'', t)\hskip 1cm X= (x, y)= (x',  x'' , y',
y'').\leqno (7.2)$$ The function  $\widetilde \psi _{\Lambda }$ is
not changed when the variables $(x', y')$ and $(x'', y'')$ are
permuted. It will be useful to make also the substitution of
$(x'_{\lambda}, y'_{\lambda})$ and $(x''_{\lambda},
y''_{\lambda})$ not everywhere, but only for $\lambda$ in a subset
of $\Lambda$. We denote by $G_{\Lambda} (E_1 , E_2)$ the set of
maps $\sigma : \lambda \rightarrow \sigma _{\lambda} $ of $\Lambda
$ in the group of substitutions $S_2$, which are constant on $E_1$
and on $E_2$, where $E_1$ and $E_2$ are disjoint subsets of
$\Lambda$. This group has a natural action $X \rightarrow \sigma
X$ on $(\R^{2p})^{\Lambda}$ and on $(\R^{4p})^{\Lambda}$. If $X$
is like in (7.2), we may have $(\sigma X) _{\lambda} =
X_{\lambda}$ or  $(\sigma X) _{\lambda} = ( x''_{\lambda} ,
x'_{\lambda}, y''_{\lambda}, y'_{\lambda})$. If $T$ is in $ {\cal
L}( \widetilde {\cal H} _{\Lambda})$, with integral kernel $K$,
let us denote by $T^{(\sigma)}$ the transformed operator, with
integral kernel $K^{(\sigma)}(X) = K(\sigma X)$. For each set $F
\subset \Lambda$, we define an element $\tau _F$ of $G_{\Lambda }
(E_1 , E_2)$ by: $$ \Big ( \tau _F (x' , x'' , y' , y'' ) \Big )
_{\lambda } = \left \{ \matrix { (x_{\lambda }'' , x_{\lambda}' ,
y_{\lambda}'' , y_{\lambda}' ) & {\rm if} & \lambda \in F \cr
(x_{\lambda }' , x_{\lambda}'' , y_{\lambda}' , y_{\lambda}'' ) &
{\rm if} & \lambda \notin F \cr } \right . \leqno (7.3)$$
 For each $\sigma \in G_{\Lambda }(E_1 , E_2)$, we denote by ${\rm sgn}_j(
\sigma )$ the signature (in $\{ -1 , 1 \}$) of $\sigma (\lambda)$
when $\lambda $ is any point of $E_j $ ($j=1, 2$), and we set
${\rm sgn }(\sigma)= {\rm sgn}_1( \sigma ){\rm sgn}_2( \sigma )$.
With these notations, we see, by easy computations, that: $${\rm
Cov}_{\Lambda , t} (A, B) = { 1\over 2 \widetilde Z_{\Lambda} (t)
} \ {\rm Tr}_{\Lambda} \ \Big (W(t)  (A' -A'') (B' - B'') \Big )
$$ where $$ W(t) = {1 \over |G |} \sum _{\sigma \in G} {\rm
sgn}(\sigma) \left ( e^{-t \widetilde H_{\Lambda } }\right )
^{(\sigma)} \hskip 1cm  G = G_{\Lambda} (E_1 , E_2) \leqno (7.4)$$
Here $  (A' - A'') (B' - B'')$ is an operator in ${\cal L}
(\widetilde {\cal H} _{E_1 \cup E_2})$, and we omit the tensor
product with the identity on $\Lambda \setminus (E_1  \cup E_2)$.
\bigskip

\bigskip
With the notations of B. Simon [22], if $E \subset \Lambda \subset
\Z^d$,  we shall use the operator of {\it partial trace} $ {\rm
Tr} _E^{\Lambda}$, from ${\cal L}_1 (\widetilde {\cal
H}_{\Lambda})$ (the space of trace norm operators) to ${\cal L}_1
( \widetilde {\cal H}_{E})$, such that, for each $W \in {\cal L}_1
(\widetilde {\cal H}_{\Lambda})$, and for each operator $T \in
{\cal L} (\widetilde {\cal H} _{E})$, we have: $${\rm Tr}
_{\Lambda} (WT) = {\rm Tr} _{E} \Big ( T \ {\rm Tr} _E^{\Lambda}
(W) \Big ).  $$ With these notations, we have: $${\rm
Cov}_{\Lambda , t} (A, B) = { 1\over 2 \widetilde Z_{\Lambda} (t)
} \ {\rm Tr}_{E_1 \cup E_2} \Big ( (A' - A'') (B' - B'') {\rm Tr}
_{E_1 \cup E_2}^{\Lambda} (W(t)) \Big ) \leqno (7.5)$$
\bigskip
For each set $E \subset \Z^d$, let us denote by $\Vert \  \Vert
^{\rm Tr} _{E}$ the trace norm of an operator in ${\cal H} _{E}$.
Then Theorem 1.3 will follow from the next proposition, (where we
write only what is needed for the points a) and b)):

\bigskip \noindent {\bf Proposition 7.1.} { \it With the notations and
hypotheses of Theorem 1.3,  if $ht \leq T_0$, $\varepsilon <
\delta < 1$, and if $t$ is smaller than some constant $ T_1
(\varepsilon, \delta)>0$, the operator $W(t)$ defined in (7.4) and
its integral kernel $W(X, t)$ satisfy: $$ \Vert {\rm Tr} _{E_1
\cup E_2}^{\Lambda} (W(t))
 \Vert ^{\rm Tr} _{E_1 \cup E_2}\  \leq \
 M( |E_1 \cup E_2 | , t, h, \varepsilon, \delta )
 \  \widetilde Z_{\Lambda} (t) \   \delta ^{ {\rm
 dist} ( E_1 , E_2) } . \leqno (7.6) $$
 $$ \int _{ (\R^p)^{\Lambda} \times  (\R^p)^{\Lambda} }
 | W( x' , x'' , x' , x'' , t) | dx' dx'' \leq \ t \ {\inf } (|E_1| ,
 |E_2| ) \ N(\varepsilon, \delta )\  \widetilde Z_{\Lambda} (t)\
  \delta ^{ {\rm
 dist} ( E_1 , E_2) }. \leqno (7.7) $$
}

\bigskip
 For the proof of this proposition, we shall use suitable decompositions of
the heat kernel $\widetilde U_{\Lambda }(x, y, t)$ of $e^{-t
\widetilde H_{\Lambda}  }$, and, therefore, of the kernel $W(x, y,
t)$. Each time a function $\varphi$ is written as a sum of terms
$\varphi  _Q$ associated, for example, to the boxes $Q$ contained
in a box $\Lambda$, the Mayer decomposition is a  simple way to
write the exponential $e^{- \varphi}$ as a sum of terms associated
to the set of boxes $\Gamma \subset {\cal P} ({\rm Box}
(\Lambda))$. This technique is often used for classical spin
systems in a lattice (see B. Simon [22]). Let us describe this
technique in our situation.

\bigskip
\noindent {\it  3. Mayer decomposition. }
 For each box $Q$, let $T_Q$ be the
operator defined by (3.7) with $\R^p$ replaced by $\R^{2p}$. We
apply this operator to the function $\widetilde \psi _{\Lambda
}(., t)$ defined in (7.2). For each box $Q \subset \Lambda$ which
is not a single point, we set $$M_Q(t) = \sup _{X\in
(\R^{4p})^{\Lambda } } (T_Q \widetilde {\psi _{\Lambda }} ) (X, t)
\hskip 1cm f_Q (X, t)\ =\ e^{ M_Q (t) - (T_Q \widetilde {\psi
_{\Lambda }}) (X, t)} \ - 1 .\leqno (7.8)$$For each box reduced to
a single point $\lambda$, we use another notation, and we set: $$
f_{ \{ \lambda \} } (X, t) \ =\ e^{ - (T_{ \{ \lambda \} }
\widetilde {\psi _{\Lambda } })(X, t) }. \leqno (7.9)$$ We denote
by $\widetilde U_{\Lambda }^{(0)}(X, t) $ the heat kernel for the
free Laplacian in $(\R^{2p})^{\Lambda}$, and we set $$\Phi _0(X,
t)= \widetilde U_{\Lambda}^{(0)}(X, t)\ e^{ - \widetilde \psi
_{\Lambda }(0, y-x , t)} e^{ - \sum _{Q \in {\rm Box} (\Lambda )}
M_Q (t)}, \leqno (7.10)$$ where ${\rm Box}(\Lambda)$ is the set of
boxes in $\Lambda$, not reduced to single points. With these
notations, by Theorem 1.2, the kernel of $ e^{-t \widetilde
H_{\Lambda}  } $ can be written $$ \widetilde U_{\Lambda }(X, t) =
\Phi _0(X, t) \ \prod _{\lambda \in \Lambda} f_{ \{ \lambda \} }
(X, t)
 \ \prod _{Q \in {\rm Box} ( \Lambda ) } (1 + f_Q (X , t)).
  $$
In order to develop the last product, we write, for each set
 of boxes $\Gamma  \in {\cal P} ({\rm Box}
(\Lambda))$: $$K_{\Gamma } (X, t) =\  \Phi  _0(X,  t)
 \prod _{\lambda \in \Lambda}  f_{ \{ \lambda \} }  (X, t)
 \prod _{Q \in \Gamma} f_Q (X, t)  . \leqno (7.11) $$
If $\Gamma$ is the empty set, the last product is $1$, by
convention.  We have $K_{\Gamma } \geq 0$. We  denote by
$T_{\Gamma }(t)$ the  operator with integral kernel $ K_{\Gamma }
(X, t) $. With these notations, we can write the heat kernel and
our mean operator $W(t)$ in the following form:
 $$ \widetilde U_{\Lambda }(X, t) =
 \sum _{\Gamma  \in {\cal P} ({\rm Box} (\Lambda))} K_{\Gamma }(X,
 t), \hskip 1cm  e^{-t \widetilde H_{\Lambda}  }  = \sum _{\Gamma \subseteq {\rm
Box} (\Lambda )} \ T_{\Gamma}(t)
 \leqno (7.12)$$
$$W(X, t)  = { 1\over |G| } \sum _{\Gamma
\subseteq  {\rm Box} (\Lambda )\atop \sigma \in G}
 ( {\rm sgn} (\sigma)) K_{\Gamma}^{(\sigma)} (X, t)
 \hskip 1cm G = G_{\Lambda}(E_1, E_2).
 \leqno (7.13)$$
 \bigskip
 Among the set of boxes appearing in the Mayer decomposition,
 some of them, called polymers, play an important role.

\bigskip

\noindent {\it  4. Polymers, and sets ${\cal C} (E_1, E_2)$ and
${\cal NC} (E_1, E_2)$.} According to the terminology of B. Simon
[22], we call {\it polymer} a finite sequence of boxes $(Q_1 , ...
,Q_k)$,  not reduced to single points, such that $Q_j \cap Q_{j+1}
\not= \emptyset $ ($1 \leq j \leq k-1)$. We say that a set $\Gamma
\in {\cal P}(B(\Lambda )) $ is in ${\cal C}  (E_1, E_2)$ if $E_1$
and $E_2$ are connected  by $\Gamma $, i.e. if $\Gamma$ contains a
polymer, with its first box intersecting $E_1$, and its last box
intersecting $E_2$. We say that $\Gamma$ is in ${\cal NC} (E_1,
E_2)$ in the opposite case.
\bigskip
 All these tools will be used together in Section 8 for the proof
 of Proposition 7.1, and therefore of Theorem 1.3.
 Since the case of two observables which are
 multiplications  is much simpler, let us
 prove the point b) of Theorem 1.3  now. The  next two lemmas will show the
role of the sets ${\cal C}(E_1 , E_2)$ and ${\cal NC}(E_1 , E_2)$.

\bigskip \noindent {\bf Lemma 7.2.}  {\it a) There exists a function
$a(t, \varepsilon)$ of the form
$a(t, \varepsilon) = t K(\varepsilon) e^{t K(\varepsilon)}$,
such that, for each polymer
 $\Pi = (Q_1 , \ldots , Q_k)$,
 we have:
$$ \prod _{j= 1}^k  |  f_{Q_j}(X, t)  |  \leq N( \Pi , \varepsilon , a(t,
\varepsilon)), \leqno (7.14)$$
where, for each $T$, we set:
 $$ N ( \Pi , \varepsilon, T) =
\prod _{j=1}^k  T \varepsilon ^{{\rm diam} (Q_j)} <Q_j>^{2d}
\hskip 1cm <Q> = 1 + {\rm diam}(Q). \leqno (7.15)$$

\smallskip
\noindent b) For each disjoint finite sets $E_1$ and $E_2$, for
each  $\delta $ such that $0 < \varepsilon < \delta < 1$, there
exists $T_1 (\varepsilon , \delta)$ and  $K(\varepsilon, \delta)$
such that, if $T < T_1 (\varepsilon , \delta)$, $$\sum _{\Pi \in
{\rm Pol} (E_1 , E_2)} N( \Pi , \varepsilon, T) \leq  T \inf (|E_1
| ,|E_2|) K(\varepsilon, \delta) \delta^{{\rm dist}(E_1, E_2)},
\leqno (7.16) $$ where ${\rm Pol}(E_1 , E_2)$ is the set of
polymers connecting $E_1$ and $E_2$, (the first box intersecting
$E_1$, and the last one $E_2$).

}

\bigskip \noindent {\it Proof.}  By (7.8) and by the
point 3 of Theorem 1.2, we can write, if $ht \leq T_0$, $$0 \leq
f_Q(X, t) \leq t K_0(\varepsilon ) \varepsilon^{{\rm diam}(Q)}
({\rm diam}(Q))^{2d} e^{tK_0(\varepsilon)},\leqno (7.17)$$ and
(7.14) follows easily. Let us prove the last point, assuming that
$|E_1| \leq |E_2|$. For each polymer $\Pi = (Q_1 , \ldots , Q_p)$,
let $L(\Pi)$ be the sum of the diameters of the boxes $Q_j$. We
remark  that the number of boxes with diameter $R\geq 1$
intersecting a given box of diameter $R_0\geq 0 $ is at most $( R
_0+1)^{2d} (R +1)^{2d} $. If we take the sum of $N(\Pi ,
\varepsilon, T)$ for all polymers starting from $E_1$, with a
given total length $L$, it follows that: $$\sum _{\Pi \in {\rm
Pol} (E_1 , \Z^d) \atop L(\Pi )= L} N(\Pi , \varepsilon , T) \leq
|E_1| \ \sum _{ R_1 + ... + R_p = L \atop p\geq 1, \  R_j \geq 1}
(1 + R_1)^{6d} ... (1+ R_p)^{6d} \varepsilon^L \ T^p.$$  We
remember that the number of ordered sequences $(R_1 , ... , R_p)$
such that $R_1 + ... + R_p = L$, and $R_j \geq 1$, is
$C_{L-1}^{p-1}$, and we set $\Phi (t)= \sup_{R>0}(1+R)^{6d} t^R$
for each $t \in ]0, 1[$ . If $0 < \varepsilon < \gamma$, it
follows that: $$\sum _{\Pi \in {\rm Pol} (E_1 , \Z^d) \atop L(\Pi
)= L} N(\Pi , \varepsilon , T) \leq |E_1| \ T\  \Phi ( \varepsilon
/ \gamma)\  \gamma ^L \ \Big ( 1 + T \Phi (\varepsilon / \gamma )
\Big )^{L-1} . \leqno (7.18)$$ We remark that, if $\Pi $ is a
polymer connecting $E_1$ and $E_2$, we have $ L(\Pi) \geq {\rm
dist} (E_1, E_2)$. If $\varepsilon < \delta < 1$, we apply (7.18)
with $\gamma = \sqrt { \varepsilon \delta}$. There exists
$T_1(\varepsilon, \delta)$ such that, if $T < T_1(\varepsilon,
\delta)$, we have $\gamma (1 + T \Phi (\varepsilon / \gamma) )
\leq \delta$. With this condition, we have: $$\sum _{\Pi \in {\rm
Pol} (E_1 , E_2)} N( \Pi , \varepsilon, T) \leq |E_1| \ T\  \Phi (
(\varepsilon / \delta)^{1/2} ) \ \sum _{ L= {\rm dist} (E_1 ,
E_2)}^{ \infty } \delta ^L.$$ The Lemma is proved.

\bigskip \noindent {\bf Lemma 7.3.} {\it  If $\Gamma \in {\cal NC} (E_1
, E_2)$, if $ ht \leq T_0$, we have $$\sum _{\sigma \in
G_{\Lambda} (E_1 , E_2)} ({\rm sgn}(\sigma)) \ K_{\Gamma}
^{(\sigma)}(X, t)= 0 \hskip 1cm  \forall  X = (x', x'' , x' , x'')
\in {\rm Diag} (\Lambda)$$where ${\rm Diag}(\Lambda)$ is the
diagonal of $(\R^p)^{\Lambda} \times (\R^p)^{\Lambda}$.  }
\bigskip

\noindent {\it Proof.}  Let $\Gamma \in {\cal NC} (E_1 , E_2)$.
For $k= 1, 2$, let us denote by $\widehat {E_k}$ the set of points
$\lambda \in \Lambda$ which are, either in $E_k$, or connected to
$E_k$ by a polymer in $\Gamma$. By Theorem 1.2, the function
$f_Q$, when it is restricted to the diagonal, depends only on the
variables $x_Q$. By (7.2) and (7.9), it is also invariant when all
the variables $(x' , y')$ and $(x'' , y'')$ are permuted. By a
combination of these remarks, for each $Q\in \Gamma $, $f_Q$,
restricted to the diagonal, is invariant by the map $\sigma _1 =
\tau _{\widehat E_1}$ defined in (7.3), and therefore $K_{\Gamma}
^{( \sigma _1)}(X, t) = K_{\Gamma}(X, t)$ if $X$ is in the
diagonal, and the lemma follows easily.

\bigskip \noindent {\it Proof of the point b) of Theorem 1.3.}
We can use the sum (7.13) for $W(X , t)$. When $X$ is in the
diagonal ($x' = y'$, $x'' = y''$), the contribution of the terms
$\Gamma \in {\cal NC} (E_1 , E_2)$ in this sum (7.13) vanishes by
Lemma 7.3. For each $\Gamma $ in ${\cal C}(E_1 , E_2)$, we can
choose a polymer $\Pi _{\Gamma} \in {\rm Pol} (E_1 , E_2)$
contained in $\Gamma$. Then, we have, by Lemma 7.2: $$\sum _{
\Gamma \in {\cal C}(E_1 , E_2)} |K_{\Gamma }(X, t)| \leq \sum
_{\Pi \in {\rm Pol} (E_1 , E_2) } N ( \Pi , \varepsilon, a(t,
\varepsilon) )
 \ \sum _{\Gamma' \in {\cal P} ({\rm Box}(\Lambda)) }
|K_{\Gamma '}(X, t)|  \leqno (7.19) $$ where $a(t, \varepsilon) =
t K_0(\varepsilon) e^{tK_0(\varepsilon)}$. Applying the positivity
of $K_{\Gamma}$ and (7.12), we obtain:

$$ \int _{ (\R^p)^{\Lambda} \times  (\R^p)^{\Lambda} }
 | W( x' , x'' , x' , x'' , t) | dx' dx''
\leq \ \widetilde Z_{\Lambda} (t)\
 \sum _{\Pi \in {\rm Pol}(E_1 , E_2) }
 N ( \Pi , \varepsilon, a(t, \varepsilon) ).$$ By the last
 point of Lemma 7.2, the estimation (7.7), and the point b) of
 Theorem 1.3 follow.

 \bigskip \noindent {\bf 8. Estimation of the quantum correlations in
 the general case. }
\bigskip
Now, we shall prove (7.6), and therefore Theorem 1.3 in the
general case. If the integral kernel of $W(t)$, (defined in
(7.4)), is denoted by $W(x, y, t)$, the integral kernel of the
partial trace $ {\rm Tr} _{E} ^{\Lambda } W(t)$ ($E =  E_1 \cup
E_2$) is $$K(x_E , y_E, t) = \int _{(\R^{2p})^{\Lambda}} W(x_E ,
x_{\Lambda \setminus E} , y_E , x_{\Lambda \setminus E}, t)
dx_{\Lambda \setminus E} \hskip 1cm E= E_1 \cup E_2. \leqno
(8.1)$$  We shall estimate the trace norm of this partial trace,
using the norm $ \Vert\  \Vert _{ m , m' , \mu }$ defined in
(5.1).

\bigskip \noindent {\bf Proposition 8.1. (Main estimate).} {\it  With
the notations and hypotheses of Proposition 7.1,  if $0<
\varepsilon < \delta < 1$, if $ht \leq T_0$, and if $t$ is smaller
than some constant $T_1 (\varepsilon, \delta)$, we have, for some
constants $m$, $m'$ and for some function  $ \mu (t)$: $$ \Vert {\rm Tr} _{E_1 \cup
E_2}^{\Lambda} (W(t))
 \Vert ^{\rm Tr} _{E_1 \cup E_2}\  \leq \
 M( |E_1 \cup E_2 | , t, h, \varepsilon, \delta )
 \     \delta ^{ {\rm
 dist} ( E_1 , E_2) }\ \sum _{\Gamma \subseteq {\rm Box}
 (\Lambda)}  \Vert K_{\Gamma} (t) \Vert _{ m , m' , \mu (t) }
 . \leqno (8.2) $$
 }
\bigskip
Classically, there exist $C>0$, $m>0$ and $m'>0$, depending on
$|E|$, such that, for each operator $K$ with integral kernel in
the Schwartz space, we have, using the norm
$ \Vert \Vert _{ m , m' , \mu }$ defined in (5.1):
$$ \Vert {\rm Tr} _{E}^{\Lambda} (K)
\Vert ^{\rm Tr} _{E}\  \leq \ C  \Vert K \Vert _{ m , m' ,
0}.\leqno (8.3) $$
 For each $\Gamma \in {\cal NC} (E_1 , E_2)$, let us denote by
  by  $H_{\Gamma}$ the subgroup of all $\sigma \in G_{\Lambda}(E_1 ,  E_2)$
  which are constant on each connected component of $\Gamma$.
  We define:
   $$K_{\Gamma}^{\rm Av}  ( t)={ 1 \over |H_{\Gamma}|}\  \sum _{\sigma \in H_{\Gamma}}
 {\rm sgn} (\sigma) K_{\Gamma}^{(\sigma)}(t), \leqno (8.4) $$
and we denote by $K_{\Gamma}^{\rm Av}  (X,  t)$ the integral
kernel. By (7.13) and (8.3), we can write: $$ \Vert {\rm Tr} _{E_1
\cup E_2}^{\Lambda} (W(t))
 \Vert ^{\rm Tr} _{E_1 \cup E_2}\  \leq \ C \Bigg [
 \sum
 _{\Gamma \in {\cal C} (E_1 , E_2)} \Vert K _{\Gamma} (t) \Vert _{ m , m' , 0}
+  \sum
 _{\Gamma \in {\cal NC} (E_1 , E_2)} \Vert K _{\Gamma}^{\rm Av} (t)
 \Vert _{ m , m' , 0} \Bigg ] . \leqno (8.5)$$
 For the first sum, we need only a modification of the estimates (7.19) and (7.16),
 because we need also bounds for the derivatives of the functions.
  Without writing all the
 details, we obtain:
 $$ \sum
 _{\Gamma \in {\cal C} (E_1 , E_2)} \Vert K _{\Gamma} (t) \Vert _{ m , m' , 0}
\leq  M( |E_1 \cup E_2 | , t, h, \varepsilon, \delta )
 \     \delta ^{ {\rm
 dist} ( E_1 , E_2) }\ \sum _{\Gamma \subseteq {\rm Box}
 (\Lambda)}  \Vert K_{\Gamma} (t) \Vert _{ m , m' , \mu (t) }
 . \leqno (8.6) $$
Now, we have to estimate  $\Vert K _{\Gamma}^{\rm Av} (t)
 \Vert _{ m , m' , 0}$ for $\Gamma \in {\cal NC} (E_1 , E_2)$.
 Therefore,
we need a study of $K_{\Gamma } ^{\rm Av} (X, t) $ only in the
following set: $${\rm Diag} (\Lambda , E_1 \cup E_2) = \{ (x, y)
\in (\R^{2p})^{ \Lambda} \times (\R^{2p})^{ \Lambda} , \ \ \ \ \
x_{\lambda} = y_{\lambda} \ \ \ \forall \lambda \notin E_1 \cup
E_2 \} . \leqno (8.7)$$
 We shall write $f\sim g$ if $f(X, t)= g(X, t)$ for all $X$ in ${\rm
Diag} (\Lambda , E_1 \cup E_2)$. Proposition 8.1 will follow from
the above inequalities and from:

\bigskip \noindent {\bf Proposition 8.2.}
{\it  For each $\Gamma$  and $\Gamma '$ such that   $\Gamma \in
{\cal NC} (E_1 , E_2)$   and $\Gamma ' \subseteq \Gamma$, we shall
find a function $B_{\Gamma , \Gamma '} (X, t)$ such that, if  $ht
\leq T_0$:
 $$ K_{\Gamma } ^{\rm Av} (X, t)
 \sim  \sum _{\Gamma  '  \subseteq \Gamma } B_{\Gamma , \Gamma '} (X, t) K_{\Gamma '} (X,
  t) \leqno (8.8)$$
For each $\Gamma '\in {\cal NC} (E_1 , E_2)$, the function
$S_{\Gamma'}$ defined by:
$$S_{\Gamma'} (X, t) = \sum _{\Gamma \in
{\cal NC} (E_1 , E_2)\atop \Gamma ' \subseteq \Gamma}
   \Big | B_{\Gamma , \Gamma '} (X, t) \Big | \leqno (8.9) $$
 satisfies, if  $0 < \varepsilon <
 \delta$, $ht \leq T_0$, if $t$ is smaller than some constant $T_1
 (\varepsilon, \delta)$, and if $X \in {\rm Diag} (\Lambda , E_1 \cup
 E_2)$, $$| S_{\Gamma'} (X, t) | \leq K( t, h, \varepsilon, \delta
 , |E_1 \cup E_2|)\ e^{\mu (t) |x_E - y_E|_1} \ \delta ^{{\rm
 dist} (E_1 , E_2)}, \leqno (8.10) $$ where $\mu (t)$ and $K $
  are some function,
 independent of $\Lambda $ containing $E_1$ and $E_2$, and depending
 on $E_1$ and $E_2$ by the number $|E_1 \cup E_2|$.
   We have similar estimates for the
 derivatives of $ B_{\Gamma , \Gamma'} (X, t)$ with respect to $x_E$ and
 $y_E$. The constants in the inequalities depend on the order of
 derivation, but not the condition of validity $t < T_1
 (\varepsilon , \delta)$.
 }
\bigskip

The rest of this section will be devoted to the proof of this
Proposition. Let us introduce some functions which are bounded like
the LHS of (8.10). Then, we shall give an expression of
$ K_{\Gamma } ^{\rm Av} (X, t)$ as a polynomial expression of such
functions.

\bigskip

   For each $\Gamma \in {\cal NC} (E_1 , E_2)$,
 we denote by $\widehat {E_k}$ ($k=1, 2$), the set of points $\lambda \in \Lambda$
wich are, either in $E_k$, or connected to $E_k$ by a polymer in
$\Gamma$.  The maps  $ \tau _{ \widehat E_k}$  defined in (7.5)
will be denoted by  $\sigma _k$.
 By applying the operators $\sigma _1$ and $\sigma  _2$ to the functions $f_Q$ of
(7.8), or to the functions $f_{\lambda}$ of (7.9),  we define the
following functions $U_Q$, $V_Q$, $W_Q$,  $U_{\lambda}$,
$V_{\lambda}$ and $W_{\lambda}$, by:
$$U_Q = f_Q^{(\sigma _1)} -
f_Q, \hskip 1cm U_{\lambda} = \left  [ f_{\lambda}^{(\sigma _1)} -
f_{\lambda} \right ] f_{\lambda}^{-1}.\leqno (8.11) $$
$$V_Q =
f_Q^{(\sigma _2)} - f_Q
 \hskip 1cm
V_{\lambda} = \Big [f_{\lambda} ^{(\tau _2)} - f_{\lambda}\Big ]
f_{\lambda}^{-1} . \leqno (8.12)$$ $$W _{Q} :=  f_{Q} -
f_{Q}^{(\sigma _1)} - f_{Q}^{(\sigma _2)}  + f_{Q}^{(\sigma _1
\sigma _2)} \hskip 1cm W _{\lambda} := \left [ f_{\lambda} -
f_{\lambda}^{(\sigma _1)} - f_{\lambda}^{(\sigma _2)}  +
f_{\lambda}^{(\sigma _1 \sigma _2)} \right ] f_{\lambda} ^{-1}
.\leqno (8.13) $$ The definition of these functions depends on the
set $\Gamma$ since $\sigma _1$ and $\sigma _2$ depend on it.  For
the estimations of these functions, we shall use the following
ones, where $X= (x, y) = (x' , x'' , y' , y'')$,  $Q$ is a box,
and $T>0$: $$M(Q , X)= <Q> ^{2d}
 \sum _{\alpha \in E_1\cup E_2}|x_{\alpha}-y_{\alpha} |
 \varepsilon^{{\rm diam } ( Q \cup \{
\alpha \} ) }, \hskip 1cm <Q> = 1 + {\rm diam}(Q)  \leqno  (8.14)$$
$$ N (Q, X, \varepsilon, T) = T \sum _{\alpha \in E_1 \atop \beta \in E_2}
|x_{\alpha}-y_{\alpha}| |x_{\beta}-y_{\beta }| \varepsilon ^{{\rm
diam} (Q \cup \{
 \alpha , \beta \} )}<Q>^{4d}. \leqno (8.15)$$

\bigskip \noindent {\bf Lemma 8.3.} {\it With these notations, we can
write, for each $\Gamma \in {\cal NC}(E_1 , E_2)$,  for each boxes
$P$ and $Q$ such that $P$ is disjoint from $\widehat E_1$ and $Q$
is disjoint from $\widehat E_2$, if $X \in {\rm Diag} (\Lambda ,
E_1 \cup E_2)$, and $ht \leq T_0$:
$$
 |U_P (X, t)| \leq a(t, \varepsilon) M(P, X , \varepsilon ) \hskip 1cm
|V_Q (X, t)| \leq a(t, \varepsilon) M(Q, X , \varepsilon ), \leqno
(8.16)$$
 where  $a(t, \varepsilon ) = t K(\varepsilon) e^{t
K(\varepsilon)}$,   ($K(\varepsilon)$ being independent of
$\Lambda$, $E_1$ and $E_2$).  If $Q$ is disjoint from $\widehat
{E_1}$ and $\widehat {E_2}$, we have: $$ |W_{Q} (X, t)| \leq N
\Big (Q, X, \varepsilon , a(t, \varepsilon) \Big ) \leqno (8.17)$$
 We have also similar estimations for points.
 If $\Phi _0$ is defined in (7.10), there exists a
 function $\Delta _0$ such that:
 $$\Phi_0^{(\sigma _1)} - \Phi _0 \sim  \Phi _0 \Delta _0
 \hskip 1cm |\Delta _0 (X, t)| \leq  t K(\varepsilon)
 \varepsilon ^{{\rm dist} (E_1 , E_2)}
   e^ { t K(\varepsilon) |x_E - y_E | _1 } \leqno (8.18)  $$
   There exists $K(\varepsilon)$
such that, for each $T>0$, for each finite set $\Lambda$,  $$\sum
_{E \subset \Lambda } \prod _{\lambda  \in E} T M (\lambda , X ,
\varepsilon ) +
\sum _{E \subset {\rm Box}(\Lambda) } \prod _{Q \in E}  T M (Q ,
X , \varepsilon ) \leq
 e^ { T K(\varepsilon) | x_E - y_E | _1 }. \leqno (8.19)  $$

}

\bigskip
\noindent {\it Proof.}   When it is restricted to
${\rm Diag} (\Lambda , E_1 \cup E_2)$, $f_P$ depends only on
$x _{E_1} - y_{E_1}$, $x _{E_2} - y_{E_2}$, and $x_P$. If
$P$ is disjoint from $\widehat E_1$, the map
$\sigma _1 = \tau _{\widehat E_1}$  has
the same effect, on $f_P$  restricted to
${\rm Diag} (\Lambda , E_1 \cup E_2)$, that the  permutation
$\tau _{E_1}$.
Then the estimation of $U_P$ follows from
Theorem 1.2.
  For the functions associated to points, we need also the last statement of
Theorem 1.2, which shows that the function $T_{ \{ \lambda \} }
\widetilde \psi _{\Lambda} (., t)$, up to an error ${\cal O}(  (t
+ h^2 t^2)$, is equal to a function which depends only on
$x_{\lambda}$ and $y_{\lambda}$, and is invariant by $\sigma _1$
and $\sigma _2$. By the form (7.10) of $\Phi _0$,   by the form
(7.2) of $\widetilde \psi _{\Lambda }$, the equality in (8.18)
will be satisfied if we choose $ \Delta _0 = e^{g} - 1$, with
$$g(X, t) = {1 \over 2} \Big [ \widetilde \psi _{\Lambda} ( 0, y -
x , t) +  \widetilde \psi _{\Lambda} ^{(\sigma _1 \sigma _2 )}( 0,
y - x , t) - \widetilde \psi _{\Lambda}^{(\sigma _1)}  ( 0, y - x
, t) - \widetilde \psi _{\Lambda} ^{(\sigma _2)} ( 0, y - x ,
t)\Big ] .$$ The inequality of (8.18) follows from Theorem 1.1 if
$ht \leq T_0$. The last inequality (8.19) is a consequence of the
following: $$ \sum _{ \lambda \in  \Z^d} M(\lambda , X ,
\varepsilon ) +   \sum _{ Q \in  {\rm Box} (\Z^d)} M(Q , X ,
\varepsilon ) \leq K(\varepsilon) |x_E - y_E | _1 .  $$ \hfill
\carre
\bigskip

Among the functions defined in  (8.11)-(8.13), only the functions
$N(Q, X , \varepsilon , T)$ and its analogue for points  have a
good rate of decay, like in (8.10),  when ${\rm dist} (E_1 , E_2)$
is large. They are used to estimate the functions $W_Q$ and
$W_{\lambda}$. Beside these functions, we need also other
functions, associated to polymers. If $\Pi = (Q_1 , ... , Q_k) $
($k\geq 2$) is a polymer, which does not connect $E_1$ and $E_2$,
we have no information on the sum of the lengths of its boxes, and
the product $f_{Q_1} \ldots f_{Q_k}$ has not necessarily the good
rate of decay. However, we shall see that, if $U_Q$ and $V_Q$ are
defined in (8.11) and (8.12),  the functions $U_{Q_1} f_{Q_2}
\ldots f_{Q_{k-1}} V_{Q_k}$ has  a good rate of decay, like in
(8.10), in terms of  ${\rm dist}(E_1 , E_2)$. In order to make
this idea more precise, let us define the functions, used for the
estimations.

\bigskip
If $\Pi = (Q_1 , ... , Q_k) $ ($k\geq 2$) is a polymer, we set, for
each points $\alpha \in E_1$ and $\beta \in E_2$, and for each
$T>0$: $$N_{\alpha , \beta }
 (\Pi ,  \varepsilon, T) = T^k
 \varepsilon^{{\rm diam }(Q_1 \cup \{ \alpha \} ) +
 {\rm diam }(Q_k \cup \{ \beta \} )}\  \prod _{j= 2}^{k-1}
  \varepsilon ^{{\rm diam } (Q_j)}\ \prod _{j=1}^k  <Q_j>^{3d}. \leqno (8.20)$$
Then we set:
$$N(\Pi  , X , \varepsilon, T) = \sum _{\alpha \in E_1 , \beta \in E_2}
  (1+ |x_{\alpha } - y_{\alpha } | ) \
  (1+ |x_{\beta } - y_{\beta } |)
  N_{\alpha , \beta } (\Pi ,  \varepsilon, T).\leqno (8.21)$$

\bigskip \noindent {\bf Lemma 8.4.} {\it  If $\Pi = (Q_1 , ... , Q_k) $
is a polymer, all its boxes belonging to $\Gamma \in {\cal NC}
(E_1 , E_2)$, we can write, with   some function $a(t,
\varepsilon)$:
\smallskip \noindent a)
If $\Pi $ is starting from $E_1$ (i.e. $Q_1 \cap E_1 \not=
\emptyset$), and if $\mu \in Q_k$, $$| f_{Q_1} ...f_{Q_{k-1}}
V_{Q_k} (X, t)  | + | f_{Q_1} ...f_{Q_{k-1}} f_{Q_k} (X, t)
V_{\mu} (X, t) | \leq N( \Pi , X , \varepsilon , a(t, \varepsilon)
). \leqno (8.22) $$ b) We have similar results if $\Pi $ starts
from $E_2$.
\smallskip \noindent c)
 If all the boxes $Q_1 , \ldots , Q_k$ of $\Pi $ are disjoint
  from $\widehat E_1$ and $\widehat E_2$,  if $\lambda \in Q_1$ and
  $\mu \in Q_k$, we can write:
  $$ | U_{Q_1} ... V_{Q_k} (X, t) | +  |U_{\lambda }f_{Q_1}  ... V_{Q_k}(X ,
  t)|  \leq
  N(\Pi , X , \varepsilon, a(t, \varepsilon)) ,   \leqno (8.23)$$
  $$    | U_{Q_1} ... f_{Q_k}  V_{\mu}(X , t)| +
  |U_{\lambda } f_{Q_1} ...f_{Q_k}   V_{\mu}(X , t) | \leq
  N(\Pi , X , \varepsilon, a(t, \varepsilon)) ,   \leqno (8.24)$$
  The factors which are not written in these products are the $f_{Q_j}$.
  In the points a), b) and c), we have similar estimations for the
  derivatives.
  \smallskip \noindent d)
We can write also, if $0< \varepsilon < \delta <1$ and $t \leq
T_1(\varepsilon, \delta)$:
 $$ \sum _{\Pi} N(\Pi, X , \varepsilon,
T) \  \leq K ( T , \varepsilon, \delta , |E| )\ \Big ( 1 + | x_E -
y_E | _1  \Big )^2 \ \delta ^{{\rm dist} (E_1 , E_2)}  \leqno
(8.25)$$where the sum is taken on all the polymers in $\Z^d$, and
$N(\Pi, X , \varepsilon, T)$ is defined in (8.21) if $\Pi $ has at
least two boxes,  and in (8.15) if $\Pi$ is reduced to a single
box $Q$. }

\bigskip

The points a), b) and c) follow from Lemma 8.2. With our
hypotheses, if $\Pi$ is starting from $E_1$, all its boxes $Q$
satisfy $Q \cap \widehat E_2 = \emptyset$, and we can apply (8.16)
for $V_{Q_k}$.  For the last point, given two boxes $Q$ and $Q'$,
we make first
 a summation on all polymers connecting $Q$ and $Q'$. Then we make a
 sum over  the boxes $Q$ and $Q'$. For the first sum, we apply Lemma
 7.2 with $E_1$ and $E_2$ replaced by $Q$ and $Q'$. For the second
 summation, we use  the following
inequality, if $0 < \varepsilon < \delta $ and $\delta _1 = \sqrt
{ \varepsilon \delta}$: $$ \sum _{ Q , Q' \in {\rm Box} (\Z^d)}
<Q> ^{4d} <Q'> ^{4d} \delta _1^{ {\rm diam} (Q \cup \{ \alpha \} )
+ {\rm diam} (Q' \cup \{ \beta \} ) + {\rm dist} (Q , Q')} \leq
C(\varepsilon , \delta) \  \delta ^{| \alpha - \beta  | }.\leqno
(8.26) $$ \hfill \carre

\bigskip \noindent {\it Proof of Proposition 8.2.
Step  1. Generators of $H_{\Gamma}$.}
 For each set of boxes $A$,
let us denote by $A_{\rm pct}$ the corresponding  set of points.
We shall denote  $\tau _A$ and $\tau _{\lambda}$ instead of $\tau
_{A_{{\rm} pct}} $  and $\tau_{ \{ \lambda \} }$ the operators
defined like in (7.3). Let ${\rm Comp} (\Gamma)$ be the set of
connected components $A$ of $\Gamma$ such that $A_{\rm pct }$ is
disjoint from $E_1 \cup E_2$, and therefore from $\widehat E_1
\cup \widehat E_2$. We set ${\rm Ext} (\Gamma) = \Lambda \setminus
( \Gamma_{\rm pct} \cup E_1 \cup E_2)$. The group $H_{\Gamma}$ is
generated by the $\tau _{A}$ ($A \in {\rm Comp} (\Gamma)$), by the
$\tau_{ \{ \lambda \} }$ ($\lambda \in {\rm Ext} (\Gamma) $), and
by the elements $\sigma _k = \tau _{\widehat E_k}$ ($k=1, 2$)
already introduced. Therefore $$ |H_{\Gamma}| = 2^{ |{\rm
Comp}(\Gamma ) | + | {\rm Ext} (\Gamma) | + 2}$$
 The generators of the group $H_{\Gamma}$, listed above, have a
  different action on the factors of the product (7.11) defining
  $K_{\Gamma}$.  Let us distinguish them.
For each connected
component $A$ in ${\rm Comp}(\Gamma)$, let: $$F_{A}= \prod _{Q\in A}
f_Q \ \prod _{\lambda \in A _{\rm pct } } f_{\lambda}.$$ For the
connected components containing a box intersecting $E_1$, or $E_2$, we
 set:
  $$\Phi _k(X, t)=  \prod _{Q\in \Gamma \atop Q_{\rm pct} \subset
  \widehat E_k }  f_Q\
  \prod _{\lambda \in \widehat {E_k} } f_{\lambda}\ \ \ \ \
  (1 \leq k \leq 2) .$$
  Then, using also the function $\Phi _0$ of (7.10),
   we can write:$$K_{\Gamma} = \Phi _0 \Phi _1 \Phi _2
    \ \prod _{A \in {\rm Comp} (\Gamma)} F_A \
    \prod _{\lambda \in {\rm Ext} (\Gamma)} f_{\lambda}
  . $$
By theorem 1.2, for each box $Q$, the function  $f_Q(., t) $,
restricted to ${\rm Diag} (\Lambda , E_1 \cup E_2)$, depends only
on the variables $x_{E_1} - y_{E_1}$, $x_{E_2} - y_{E_2}$, $x_{Q}$
and $y_Q$. The equality
(7.2) shows that this function is invariant when all the variables
$(x' , y')$ and $(x'' , y'')$ are permuted. From these two
remarks, some properties follow for the functions defined above.
We can write, if $A$ and $B$
are in ${\rm Comp} (\Gamma )$, $A \not= B$, if $\lambda$ and $\mu$
are in ${\rm Ext}(\Gamma)$, $\lambda \not= \mu$ : $$ F_A ^{ (\tau
_B) } \sim F_A \ \ \, \hskip 1cm F_A ^{ (\tau _A) } \sim
F_A^{(\sigma _1 \sigma _2)} \hskip 1cm
 F_A ^{ (\tau _{\lambda}) } \sim F_A \hskip 1cm f_{\lambda}^{(\tau _A)}
 \sim f_{\lambda} \leqno (8.27)$$
 $$f_{\lambda } ^{(\tau_{\mu})}\sim f_{\lambda }  \hskip 1cm
 \Phi _k ^{(\tau _A) } \sim  \Phi _k ^{(\tau _{\lambda} ) }\sim
 \Phi _k \ \ \ \ (0 \leq k \leq 2)
 \hskip 1cm \Phi _1^{(\sigma _1)} \sim \Phi _1^{(\sigma _2)}, etc\ldots $$
 Let us denote by $H_{\Gamma}^+$ the subgroup  of
$ H_{\Gamma}$ generated by the $\tau _A$ ($A \in {\rm Comp}
(\Gamma)$) and $\tau _{\lambda } $ ($\lambda \in {\rm
Ext}(\Gamma)$). By the above equalities and similar ones, we can
write:
 $$K_{\Gamma} ^{\rm Av}  \sim 2
K_{\Gamma}^+ - 2 \Big ( K_{\Gamma}^+ \Big )^{(\sigma _1)} ,
 \hskip 1cm K _{\Gamma}^+ (X, t)= { 1 \over  | H_{\Gamma}  | }
 \sum _{\sigma \in H_{\Gamma}^+ }
 K_{\Gamma}^{(\sigma)} .\leqno (8.28) $$
 We have also:
$$K_{\Gamma}^+
\sim   { 1 \over  | H_{\Gamma}  | }  \Phi _0 \Phi _1 \Phi _2 \
  \prod _{ A\in {\rm Comp}  (\Gamma)}
 \Big [  F_{A} + F_{A}^{(\sigma _1 \sigma _2)}\Big ]\
 \prod _{\lambda \in {\rm Ext}(\Gamma)} \Big [ f_{\lambda} +
f_{\lambda}^{(\sigma _1 \sigma _2)} \Big ], \leqno (8.29) $$
$$(K_{\Gamma }^+)^{(\sigma _1 )} \sim   { 1 \over  | H_ {\Gamma} |
}       \Phi _0^{(\sigma _1)} \Phi _1^{(\sigma _2)}  \Phi
_2^{(\sigma _1)} \ \prod _{ A\in {\rm Comp} (\Gamma)}
 \Big [  F_{A}^{(\sigma _1)} + F_{A}^{(\sigma _2)}\Big ]\
 \prod _{\lambda \in {\rm Ext}(\Gamma)} \Big [ f_{\lambda}^{(\sigma _1)} +
f_{\lambda}^{( \sigma _2)} \Big ]\leqno (8.30) $$

\medskip \noindent {\it  Step 2. Polynomial expression of $K_{\Gamma}^{\rm
Av} $.} Now, we shall write the difference between (8.29) and
(8.30) as a polynomial expression of functions that are bounded
like in (8.10). We remember that: $$f_Q^{(\sigma _1)} = f_Q +
U_Q,\hskip 1cm f_Q^{(\sigma _2)} = f_Q + V_Q \hskip 1cm
f_Q^{(\sigma _1 \sigma _2)} = f_Q + U_Q + V_Q + W _Q .$$According
to the notations (8.11)-(8.18), we have $f_{\lambda}^{(\sigma
_1)}= f_{\lambda} (1 + U_{\lambda})$, etc... and $ \Phi
_0^{(\sigma _1)}  \sim  (1 + \Delta _0 ) \Phi _0$. Thus we can
write $$K_{\Gamma} ^{\rm Av} \sim
 G_{\Gamma} \
 \Phi _0 \  \prod _{\lambda \in \Lambda } f_{\lambda}
 ,  $$
 where $G _{\Gamma} $ is a polynomial expression of the functions $f_Q$, $U_Q$, $V_Q$ and
$W_Q$ ($Q\in {\rm Box} (\Lambda)$), of the functions $U_{\lambda}$,
$V_{\lambda}$ and $W_{\lambda}$ ($\lambda \in \Lambda$), and of
the function $\Delta _0$. Let us describe more carefully this
polynomial. Let $I$ be the set of partitions of $\Gamma$ in four
subsets ${\cal F}$, ${\cal U}$, ${\cal V}$, ${\cal W}$. Let  $I_p$ be the
set of triples  $({\cal U}^p, {\cal V}^p, {\cal W}^p)$
such that
${\cal U}^p$, ${\cal V}^p$, ${\cal W}^p$ are disjoint subsets
of $\Lambda$.  Let $J = I \times I_p \times \{ 0, 1 \}$.
Each element $j\in J$ will be written
 $j = ({\cal F}_j , .... , {\cal W} _j^p , m_j)$.
Thus we can write:
 $$K _{\Gamma} ^{\rm Av}  \sim  \sum
_{j \in J} c_j G _{\Gamma}  ^{ [j]}
\hskip 1cm G _{\Gamma}  ^{ [j]}  =
 \Phi _0 \
\Delta _0^{m_j} \ \prod _{Q \in {\cal F}_j} f_Q \ \prod _{Q \in
{\cal U}_j} U_Q ... \prod _{\lambda \in {\cal W}_j^p } W_{\lambda}
 \ \Big [  \prod _{\lambda \in \Lambda } f_{\lambda}\Big ]
\leqno (8.31) $$ where the coefficient $c_j$ are constant, and
uniformly bounded. Moreover, we have $c_j= 0$ unless one of the
following three conditions $(A)$, $(B)$ or $(C)$ is satisfied. $$
{\cal W}_j \not= \emptyset \ \ \ \ {\rm or } \ \ \ \ \ {\cal
W}_j^p \not= \emptyset \ \ \ \ {\rm or } \ \ \ \ m_j=1 \leqno
(A)$$ $$ ({\cal V}_j)_{\rm pct} \cap \widehat {E_1} \not=
\emptyset\ \ \ \ \ {\rm or} \ \ \ \ {\cal V}_j^p \cap \widehat
{E_1} \not= \emptyset \ \ \ \ \ {\rm or}\ \ \ \ ({\cal U}_j) _{\rm
pct}  \cap  \widehat {E_2} \not=  \emptyset\ \ \ \ \ {\rm or} \ \
\ \ {\cal U}_j^p \cap \widehat {E_2} \not= \emptyset \leqno (B)$$
\noindent (C) There exists a connected component $A \in {\rm Comp
}(\Gamma)$ such that, in one hand, $A \cap {\cal U}_j \not=
\emptyset$, or $A_{\rm pct} \cap {\cal U}_j^p \not= \emptyset$,
and in the other hand, $A \cap {\cal V}_j \not= \emptyset$, or
$A_{\rm pct} \cap {\cal V}_j^p \not= \emptyset$.

\smallskip
Moreover, we have also $c_j=0$ unless all the following condition
are satisfied, and also the similar ones for the sets of points:
$$({\cal U}_j)_{\rm pct}  \cap \widehat {E_1}  = \emptyset \hskip
1cm ({\cal V}_j ) _{\rm pct} \cap \widehat {E_2} = \emptyset
\hskip 1cm ({\cal W}_j ) _{\rm pct} \cap ( \widehat {E_1}  \cup
\widehat {E_2}  ) = \emptyset .   $$

\medskip \noindent {\it Step 3. Construction of $B_{\Gamma , \Gamma '}$.}
For each $j\in J$ such that $c_j \not= 0$, we shall write the term
$G _{\Gamma}  ^{ [j]}$ in (8.31) in the form: $$ G _{\Gamma}  ^{
[j]} = B_j K_{\Gamma _j} \leqno (8.32) $$ where $K_{\Gamma _j}$ is
defined as in (7.11), with $\Gamma $ replaced by some subset
$\Gamma _j$. The function $B_j$ will have the rate of decay of
(8.10) when ${\rm dist} (E_1 , E_2)$ is large, because the product
defining $B_j$ will contain, either $\Delta _0$, or a function
$W_{\lambda}$, or a function, like those of Lemma 8.4,
corresponding to a polymer $\Pi _j$. In order to apply Lemma 8.4,
we shall need sometimes one point $\lambda _j$, or two. We shall
denote by ${\cal X} _j$ the set of points needed for the
application of Lemma 8.4: this set has $0$, $1$ or $2$ points. Let
us define $\Gamma_j$, $\Pi _j$ and ${\cal X}_j$ in all the cases
A, B and C.
\medskip
If $j$ satisfies (A), let $\Gamma _j = {\cal F}_j$, and let $B_j$
be defined by (8.32). In the first  case of (A), the  polymer $\Pi
_j$ is reduced to single box $Q$ chosen in $ {\cal W}_j$, and
${\cal X}_j = \emptyset$.  In the second case,  $\Pi_j=
\emptyset$, and for ${\cal X}_j$, we choose one point in ${\cal
W}_j^p$. In the third case, $\Pi_j = \emptyset$, and ${\cal X}_j=
\emptyset$.
\medskip
If $j$ satisfies the first condition of (B), let $Q$ be a box in
${\cal V}_j$  such that $Q \cap \widehat E_1 \not=0$. There is a
polymer $\Pi _j$ in $\Gamma$, starting in $E_1$, the last box of
which being $Q$. Let $\Gamma _j= {\cal F}_j \setminus \Pi _j$ be
the set of the boxes in ${\cal F}_j$, excepted those we took from
${\cal F}_j$ to construct the polymer $\Pi _j$. Let $B_j$ be
defined by (8.32), and ${\cal X}_j = \emptyset$. If ${\cal V}_j^p
\cap \widehat E_1 \not= \emptyset$, let $\mu _j$ be a point in
this set. There is a polymer $\Pi _j$, starting from $E_1$, such
that $\mu _j$ is in its last box. We define $\Gamma_j$ and $B_j$
as before, but ${\cal X}_j = \{ \mu_j \}$. We proceed in the same
way in the other cases of (B).
\medskip
Now, let $j$ satisfying one of the conditions of (C), for example
such that, for some connected component $A\in {\rm Comp}(\Gamma)$,
we have $A \cap {\cal U}_j \not= \emptyset$ and $A_{\rm pct} \cap
{\cal V}_j^p \not= \emptyset$. Let $P$ is a box in the first set,
and $\mu _j$ be a point in the second one. Since they are in the
same connected component, there is a polymer $\Pi _j$, whose first
box is $P$, and such that $\mu _j$ is in its last box.  We define
$\Gamma_j$ and $B_j$ as before, and ${\cal X}_j = \{ \mu_j \}$.

\medskip
Now for all $j$, we have defined $\Gamma _j $ and $B_j$ such that
(8.32) is satisfied. If, for some $j\in J$, we are in several
cases, we make a choice.  For each $\Gamma ' \subseteq \Gamma$, we
set $$B_{\Gamma , \Gamma '} = \sum _{j \in J \atop \Gamma _j =
\Gamma'} c_j  B_j \leqno(8.33) $$ and (8.8) is satisfied.

\medskip \noindent {\it Step 4. Estimation of $B_{\Gamma , \Gamma '}$.}
When a polymer $\Pi_j$ is used for the application of Lemma 8.4,
with a set ${\cal X}_j \subset \Lambda$ with $0$, $1$ or $2$
points, we can write, by Lemmas 8.3 and 8.4, : $$ |B_j(X, t)| \leq
N( \Pi _j , X , \varepsilon, a(t , \varepsilon) ) \
 \prod _{ Q \in ({\cal U}_j
\cup {\cal V}_j\cup  {\cal W}_j)\setminus \Pi_j} a(t ,
\varepsilon) M(Q
 , X , \varepsilon )\
 \prod _{ \lambda  \in ({\cal U}_j^p
\cup {\cal V}_j^p  \cup  {\cal W}_j^p )\setminus {\cal X}_j} a(t ,
\varepsilon) M(\lambda
 , X , \varepsilon )\
.$$  In order to estimate $B_{\Gamma , \Gamma'}$, following
(8.33), we make first a summation on all the sets ${\cal X}_j$
corresponding to a same polymer $\Pi_j$: this gives only a change
in the power of $<Q>$ in the definition of $N(\Pi _j , ... )$.
Then we sum on all the sets ${\cal U}_j^p$, ${\cal V}_j^p$ and
${\cal W}_j^p$, applying the last point of Lemma 8.3. We obtain,
proceeding in a similar way for the terms without polymer:
$$|B_{\Gamma , \Gamma'} (X , t)|\ \leq \ e^{ T |x_E - y_E| _1 }
\sum _{ \Pi \in {\rm Pol}_{\Gamma \setminus \Gamma'} } \left [
N(\Pi , X , \varepsilon, T) \ \prod _{ Q \in \Gamma \setminus
(\Gamma' \cup \Pi)} T M_Q (X , \varepsilon) \right ]\hskip 1cm T=
a(t , \varepsilon) .$$ Therefore, applying the last points of
Lemmas 8.3 and 8.4, we see that  the function $S_{ \Gamma'} $
defined in (8.9) satisfies (8.10). We shall not write the details
for the derivatives of $B_{\Gamma , \Gamma '}$.

\bigskip \noindent {\it  End of the proof of Proposition 8.1.} It
remains to estimate the second sum in (8.5). By Proposition 8.2,
with the similar estimates for derivatives, we can write, for some
function $\mu _1(t)$: $$ \sum
 _{\Gamma \in {\cal NC} (E_1 , E_2)} \Vert K _{\Gamma}^{\rm Av} (t)
 \Vert _{ m , m' , 0}   \leq
 K( t, h, \varepsilon , \delta , |E|)  \delta ^{{\rm dist} (E_1 ,
 E_2)} \
  \sum
 _{\Gamma \in {\cal NC} (E_1 , E_2)} \Vert K _{\Gamma} (t)
 \Vert _{ m _1, m' _1, \mu _1(t) }.$$
If we look at the derivatives of the functions $K_{\Gamma}(X, t)$
defined in (7.11), we can write:
 $$ \sum _{\Gamma \subseteq {\rm Box}
 (\Lambda)}  \Vert K_{\Gamma} (t) \Vert _{ m _1, m'_1 , \mu _1(t) }
 \leq   M( |E| , t, h, \varepsilon )
 \sum _{\Gamma \subseteq {\rm Box}
 (\Lambda)}  \Vert K_{\Gamma} (t) \Vert _{ 0 , m'_2 , \mu_2 (t) }$$
By the positivity of $K_{\Gamma} (X, t)$ and by the equality (7.12),
we can write: $$ \sum _{\Gamma \subseteq {\rm Box}
 (\Lambda)}  \Vert K_{\Gamma} (t) \Vert _{ 0 , m'_2 , \mu_2(t) }
 \leq   \Vert \widetilde U _{\Lambda} (t)  \Vert _{ 0 , m'_2 , \mu_2 (t) } $$
 By Proposition 5.1, we can write, with
some constant $M( |E| , t, h, \varepsilon )$,
$$ \Vert
\widetilde U _{\Lambda} (t)  \Vert _{ 0 , m'_2, \mu_2 (t)} \leq M(
|E| , t, h, \varepsilon ) \ \widetilde Z_{\Lambda }
(t).$$
The main estimate of Proposition 8.1, and therefore
the estimation (7.6) of Proposition
7.1, and therefore the point a) of Theorem 1.1, are proven.
The points b) and d) were proved in section 7, and the point c),
(estimation of $K_{{\rm op} , {\rm fc}} (E_1 , E_2 , t, h)$),
is an intermediate between the
two proofs: we need a partial trace ${\rm Tr}_{E_1}^{\Lambda}$
instead of ${\rm Tr}_{E_1\cup E_2 }^{\Lambda}$. Thus we
obtain, in the estimation, a constant depending on $ | E_1 | $ instead
of $ |  E_1 \cup E_2|  $. Theorem 1.3 is proved.

  \vskip 3cm
 \noindent
 {\bf 9. Proof of Theorems 1.4 and 1.5.}
  \bigskip
  For the proof of Theorem 1.4, we consider a box $\Lambda $ in $\Z^d$,
  which is the union of two disjoint boxes $\Lambda _1$ and $\Lambda _2$,
  separated by an hyperplane $\Sigma$, like in (6.1).
  We consider a local observable $A $, supported in
  one of these sets, for example $\Lambda _1$.
  We shall estimate the difference between
  $E_{\Lambda  , t} (A)$, (the mean value of $A$, defined by (1.2),
  when $A$ is seen as an operator in ${\cal H}_{\Lambda }$),
  and  $E_{\Lambda _1 , t} (A)$, (the analogue for $\Lambda _1$).
 Theorem 1.4 will be an easy consequence of the next Proposition,
 since $\Lambda _{m+n}$ is obtain from $\Lambda _n$ (defined in
 (1.19)) by applying
 $2d$ times this procedure of enlarging.

\bigskip
\noindent {\bf Proposition 9.1.} {\it  With the above notations,
if the interaction satisfies $(H_{\varepsilon})$, if $\varepsilon
< \delta < 1$, there exists $T_1 (\varepsilon, \delta)$ and a
function $ K(t, h,  \varepsilon, \delta , N)$ such that, if $ht <
T_0$ and $t < T_1 (\varepsilon, \delta)$: $$ \Big | E_{\Lambda , t
} (A) - E_{\Lambda _1, t } (A) \Big | \leq   K(t, h,  \varepsilon,
\delta , |{\rm supp} (A) |  )\ \ \delta^{ {\rm dist}  ( {\rm supp}
(A), \Sigma ) } \Vert A \Vert . \leqno (9.1)$$

}
\bigskip

  \noindent {\it Proof. First step. } For each $\theta \in [0, 1]$,
  we shall use the potential
$V_{\Lambda , \theta} = V _{\Lambda} - \theta V_{\rm inter}$,
(where $V_{\rm inter}$ is defined in (6.2)), the corresponding
Hamiltonian $H_{\Lambda , \theta}$, the  heat kernel $U_{\Lambda ,
\theta}$, the  function   $\psi_{\Lambda , \theta}$ of  Theorem
1.1. and the corresponding correlation ${\rm Cov}_{\Lambda , t ,
\theta}$ of two operators, and the mean value  $E_{\Lambda , t ,
\theta}(A)$.  We denote by $Z_{\Lambda ,
 \theta}(t)$ the trace of the operator $e^{-t H_{\Lambda ,
 \theta}}$.  If $A$ is supported in $\Lambda _1$, we have, with these
 notations:
 $$ E_{\Lambda , t, 0 } (A) =  E_{\Lambda , t }
(A)  \hskip 1cm  E_{\Lambda , t, 1 } (A) = E_{\Lambda _1, t } (A)
\leqno (9.2) $$
 We  set,  for each $\theta \in [0, 1]$,
  $$R_{\theta} ( x, y, t) =
\partial _{ \theta}   \psi _{\Lambda , \theta} (x, y, t)
   \ - \
\partial _{ \theta}   \psi _{\Lambda , \theta} (x, x, t),
 \hskip 1cm \varphi _{\theta} (x, t)= \psi_{\Lambda ,
 \theta} (x, x, t), \leqno (9.3)$$
For a given $K(x, y)$, in $ {\cal S} ((\R
^p)^{\Lambda } \times (\R ^p)^{\Lambda })$, let $Op(K)$ be the
operator with integral kernel $K(x, y)$.  We denote by
${\rm Cov} _{\Lambda , t , \theta} (A , f)$ the correlation between an
operator $A$ and the multiplication by a function $f$.
We use
the operators $T_Q$ of Section 3, with the first definition (3.2),
 applied to  functions depending only on $x$. With all these
 notations, we have:
  $$ \partial _{\theta} E_{\Lambda , t, \theta } (A) =
\ Z_{\Lambda , \theta} (t)^{-1}
  {\rm Tr} \Big (Op \big (
 U_{\Lambda , \theta } (. , t) R _{\theta} (., t)
 \big ) \circ A \Big )\
 +\ {1\over 2} \sum _{Q \subseteq \Lambda}
{\rm Cov}_{\Lambda, t,\theta}  (A, T_Q\partial _{\theta} \varphi
_{\theta} (. , t) ). \leqno (9.4) $$

\medskip  \noindent {\it Second step.}
Using (8.3), with the norm defined in (5.1), we can write, with
$E = {\rm supp}(A)$:
$$ \Big |  {\rm Tr}(Op( U_{\Lambda , \theta }(., t)  R_{\theta } ) \circ
A ) \Big | \leq K ( |E|)\ \Vert A \Vert \
\Vert Op ( U_{\Lambda , \theta }(.,
t)  R_{\theta } ) \Vert _{ m, m' , 0}
$$ with  $m$ and $m'$ depending on $|E|$.
 By the point (6.4) of Lemma 6.2, we can write, for each points
  $\lambda ^{(1)}$, ...
$\lambda ^{(m)}$ of $E$, for each $(x, y)$ $$|\nabla _{\lambda
^{(1)}}... \nabla _{\lambda ^{(m)}} R_{\theta } (x, y, t)| \leq t
K(\varepsilon)  \ \varepsilon ^{ {\rm dist} (E, \Sigma )}.$$
Therefore,
$$\Vert Op ( U_{\Lambda , \theta }(.,
t)  R_{\theta } ) \Vert _{ m, m' , 0} \leq \
t K(\varepsilon)  \  \varepsilon ^{ {\rm dist} (E, \Sigma )}
 \Vert  U_{\Lambda , \theta }(  . , t ) \Vert _{m , m'+1 , 0} .$$
 By Proposition 5.1, applied  to the subset $E$, we can write:
$$ \Vert  U_{\Lambda , \theta }(  . , t ) \Vert _{m , m'+1 , 0}
\leq K(t, h,  \varepsilon, | E|  )  Z _{\Lambda  ,
\theta}(t),$$
Therefore:
$$|  {\rm Tr}(Op( U_{\Lambda , \theta }(., t)  R_{\theta } ) \circ
A ) | \leq  \ K(t, h,  \varepsilon, | E|  )\ \Vert A \Vert \
 Z _{\Lambda  ,\theta}(t) \   \varepsilon ^{ {\rm dist} (E, \Sigma )}
 \leqno (9.5) $$
Now, we shall estimate the second term in (9.4). By Proposition
6.1, we can write: $$ \Vert T_Q\partial _{\theta} \varphi
_{\theta} (. , t)  \Vert   \leq t K (\varepsilon) <Q> ^d
\varepsilon^{ {\rm diam} (Q \cup \pi _{ \Sigma } (Q) )  }  ,
\leqno (9.6)$$ where $\pi _{ \Sigma } (Q) $ is the orthogonal
projection of $Q$ on the hyperplane $\Sigma$, which separates
$\Lambda _1$ and $\Lambda _2$. If $\varepsilon < \delta < 1$, we
can apply Theorem 1.3, point c, for the correlation between an
operator and a function, with $\delta$ replaced by $\delta _1 =
\sqrt { \varepsilon \delta}$. Thus we can write: $$\Big | {\rm
Cov} _{\Lambda , t , \theta }  (B_Q,A) \Big | \leq K (t, h,
\varepsilon, \delta ,  | E| )\
 \Vert A \Vert \ \Vert T_Q\partial _{\theta} \varphi
_{\theta} (. , t)  \Vert  \ \delta _1^{ {\rm dist} (E , Q)}.
\leqno (9.7) $$ With the same relations between $\varepsilon$,
$\delta _1$ and $\delta$, we have: $$\sum _{ Q \subset \Z^d}
\delta _1 ^{ {\rm dist} (E , Q) + {\rm diam} (Q \cup \pi _{\Sigma
} (Q)) } \leq K(\varepsilon , \delta) \delta^{ {\rm dist} (E ,
\Sigma)}.\leqno (9.8) $$ By (9.5)-(9.8), Proposition 9.1 is
proved, and Theorem 1.4 follows easily.

\hfill  \carre

\bigskip

 Theorem 1.5, about the mean energy per site, will follow from the next Proposition.
 The mean energy
 $X_{\Lambda} (t) $ for the set $\Lambda$ is defined in (1.22).

\bigskip
\noindent {\bf Proposition 9.2.} {\it
For each box $\Lambda $ of $\Z^d$,  split
into two boxes $\Lambda _1$ and $\Lambda _2$,
separated by an hyperplane $\Sigma$ as in (6.1),
for each $t>0$ and  $h>0$ such that $ht < T_0$ and $t$ is small enough,
$$| X_\Lambda (t)\ -
 X_{\Lambda _1}(t)\ -X_{\Lambda_2}(t)\ |
 \leq K(t) \  | \Lambda_\bot|    .\leqno(9.9 ) $$
where $ \Lambda_\bot = \pi _{\Sigma } (\Lambda)$, and $\pi _{\Sigma }$
is the orthogonal projection on $\Sigma$.
}
\bigskip
\noindent {\it Proof.} For each $\theta \in [0, 1]$, let $
X_{\Lambda , \theta}(t)$ be the mean energy, for the set
$\Lambda$, but with $V_{\Lambda}$ replaced by the potential
$V_{\Lambda , \theta} = V_{\Lambda} - \theta V_{\rm inter}$, where
$V_{\rm inter}$ is defined in (6.2).
  Thus,
$X_{\Lambda , 0}(t) = X_{\Lambda } (t)$ and $X_{\Lambda , 1 }(t)=
X_{\Lambda _1} (t) + X_{\Lambda _2} (t)$. If $\psi_{\Lambda ,
\theta}(x,y,t)$ is the function of Theorem 1.1, associated to the
potential $V_{\Lambda , \theta}$, we set $\varphi _{\theta }(x,
t)= \psi_ {\Lambda , \theta }(x, x, t)$. By computations, similar
to those of Proposition 9.1, we find that $$\partial _{\theta}
X_{\Lambda , \theta}(t)= E_{ \Lambda , t , \theta} \left (
{\partial^2\phi_\theta (. , t) \over
\partial t \partial \theta
} \right ) + {1 \over 2} {\rm Cov} _{\Lambda , t , \theta} (
{\partial \phi_\theta\over\partial t} , {\partial
\phi_\theta\over\partial \theta}) = F_1(\theta , t)+F_2(\theta ,
t).\leqno (9.10)$$ where  ${\rm Cov} _{\Lambda , t , \theta}  (f,
g)$ is the correlation of the multiplications $M_f$ and $M_g$,
defined as in (1.22) for the Hamiltonian $H_{\Lambda , \theta}$,
and $E _{\Lambda , t , \theta}  (f)$ is defined in the same way.
\medskip

\noindent {\it Estimation of $F_1(\theta , t)$.} Following (2.1),
with    $V_{\Lambda}$ replaced by $V_{\Lambda , \theta }$, we
have: $${\partial \phi _{\theta}  (x, t ) \over\partial t}
=
{h^2\over 2} (\Delta _x \psi_{\Lambda , \theta } )( x, x, t)
-{h^2\over 2}\vert (\nabla _x \psi_{\Lambda , \theta } (x, x, t)
\vert^2 +V_{\Lambda , \theta } (x). $$ By differentiating with
respect to $\theta$, we obtain $$ |
\partial_{t}\partial_\theta \phi_\theta(x, t) | \leq {h^2 \over 2}
\sum_{\lambda\in\Lambda} | \Big
(\Delta_{{x_\lambda}}\partial_\theta \psi_\theta\Big ) (x, x, t)|
+ h^2 \sum_{\lambda\in\Lambda} |  \nabla
_{{x_\lambda}}\partial_\theta \psi_\theta (x, x, t) |  \
 |  \nabla _{{x_\lambda}}\psi_\theta (x, x, t)|
 +|    \partial_\theta V_{\Lambda , \theta  }   |  . $$
We remark that $$  |   \partial_\theta V_{\Lambda ,\theta } (x)  |
= | V_{\rm Inter} (x)  | \leq K  \sum _{\lambda \in \Lambda _1,
\mu \in \Lambda _2} \varepsilon ^{|\lambda - \mu |} \leq \ K
(\varepsilon) | \Lambda_\bot  |. $$ By Lemma 6.2 for $m=1$ and
$m=2$,   we can write, if  $ht \leq T_0$, $$| \nabla
_{{x_\lambda}}\partial_\theta \psi_\theta (x, x, t)  | + | \Big
(\Delta_{{x_\lambda}}\partial_\theta \psi_\theta\Big ) (x, x, t)|
\leq t K(\varepsilon) \varepsilon ^{{\rm dist } (\lambda ,
\Sigma)} ,$$ and therefore, $$ | \partial_{t}\partial_\theta
\phi_\theta(x, t) |  \leq K(\varepsilon)  h^2 (t + t^2) \
\sum_{\lambda \in \Lambda} \varepsilon ^{ {\rm dist} (\lambda ,
\Sigma)} \leq K(\varepsilon)  h^2 (t + t^2) |  \Lambda_\bot | . $$
and therefore   $|  F_1(\theta , t) | \leq
   K(\varepsilon)  h^2 (t + t^2) |  \Lambda_\bot | $.
\medskip
\noindent {\it Estimation of $F_2(\theta , t )$.} Now, we use the
decomposition of Theorem 1.2, (definition (3.7)),  for
$\psi_{\Lambda ,\theta }$. We apply (1.13) with $\psi_{\Lambda }$
replaced by this function. Now, we take the derivatives of this
equality with respect to the variables $x_{\lambda }$ ($\lambda
\in \Lambda $), we restrict then to the diagonal, and we report
into (9.17), we  take also the derivative with respect to
$\theta$, and we report both in the expression of $F_2(\theta)$
defined in (9.10). We remark that, if a function does not depend
on $x$, its correlation with any other one vanishes. In order to
write more shortly what we obtain, we introduce the following
notations: $$ v_{\lambda } (t) =  - ( \nabla  _{x_{\lambda }} \psi
 _{\Lambda , \theta } ) (0, 0 , t)\ \ \ \ \ \ \
f_{ \lambda , Q }(x, t) = (\Delta _{x_{\lambda }} T_Q \psi
_{\Lambda , \theta } ) (x, x, t) \ \ \ \ \ \ \ \  g_{ \lambda , Q }(x, t) =
(\nabla  _{x_{\lambda }} T_Q \psi _{\Lambda , \theta } )
 (x, x, t)$$
 $$h_Q(x, t) = (T_Q {\partial\over
 \partial\theta}\psi_{\Lambda , \theta }) (x, x, t).$$
 Thus, we obtain:
 $$F_2(\theta ) = {h^2 \over 4} \sum _{ \lambda \in \Lambda \atop Q,
 Q' \subseteq \Lambda} {\rm Cov}_{\Lambda , t, \theta}
 ( f_{ \lambda , Q }, h_{Q'})\
 - h^2   \sum _{ \lambda \in \Lambda \atop Q,
 Q' \subseteq \Lambda} v_{\lambda }(t)\  .\
 {\rm Cov}_{\Lambda , t, \theta}  ( g_{\lambda , Q }\ , \
 h_{Q'}) - \leqno (9.11)$$
 $$- {h^2 \over 2 } \sum _{ \lambda \in \Lambda \atop Q,
 Q', Q'' \subseteq \Lambda}  {\rm Cov}_{\Lambda , t, \theta}
 \Big ( g_{\lambda , Q}.g_{\lambda ,
 Q'}\ , \
 h_{Q''} \Big ) + \sum _{  Q, Q' \subseteq \Lambda}
 {\rm Cov} _{\Lambda , t, \theta} ( T_Q V _{\Lambda , \theta}, h_{Q'}).$$
 Now, we shall estimate all the terms.
 By Theorem 1.2, $f_{\lambda , Q }$, $g_{\lambda , Q }$
and  $h_Q$ depend only on $x_Q$ and,
 if $ht \leq T_0 $:
 $$ | f_{\lambda , Q} (x, t)|
 + |  g_{\lambda , Q} (x, t) |
  \leq t K(\varepsilon) <Q>^{2d}
  \varepsilon ^{ {\rm diam} (Q \cup \{ \lambda \} )}  $$
  By Theorem 1.1, we can write $|  v_{\lambda }(t)| \leq Ct$.
  By Proposition 6.1,  we can write:
$$|h_{Q}(x,t)|  \leq  t K(\varepsilon) <Q>^{2d} \varepsilon ^{
{\rm diam} (Q \cup \pi _{\Sigma}(Q)) }.  $$
   We apply Theorem 1.3, in the multiplicative case,
   to estimate the four correlations in the
  expression of $F_2(\theta )$. If $\varepsilon < \delta <
  1$, we apply  theorem 1.3 with $\delta$ replaced by
   $\delta _1 = \sqrt { \varepsilon \delta}$. We obtain:
 $${h^2 \over 4} | {\rm Cov}_{\Lambda , t, \theta}
 ( f_{\lambda , Q} , h_{Q'})| \leq
   h^2 t^3 K(\varepsilon, \delta)
 \delta _1^{ {\rm diam} (Q \cup \{ \lambda \} )
 + {\rm dist} (Q , Q') + {\rm diam } (Q' \cup \pi _{\Sigma}(Q'))}$$
Therefore, using again (8.26), as in the proof of Lemma 8.4, we
get: $${h^2 \over 4} \sum _{ \lambda \in \Lambda \atop Q,
 Q' \subseteq \Lambda}  | {\rm Cov} _{\Lambda , t, \theta}
 ( f_{ \lambda , Q }, h_{Q'}) |
 \leq  h^2 t^3 K(\varepsilon , \delta) \sum _{\lambda \in \Lambda
 , \mu \in \Lambda _{\bot}} \delta ^{ {\rm dist} (\lambda , \mu)}
 \leq h^2 t^3 K(\varepsilon , \delta)  | \Lambda _{\bot }|.$$
 The other terms in the expression (9.11) of $F_2(\theta)$ are bounded
 in the same way, excepted
 the terms $ {\rm Cov} _{\Lambda , t, \theta}  (T_Q V _{\theta}, h_{Q'})$
 such that $Q$ is reduced to
 a single point $\lambda$,
 for which we need the point d) of Theorem 1.3:
 $$|{\rm  Cov } _{\Lambda , t, \theta}
 (T_{ \{ \lambda \} }  V _{\theta}, h_{Q'})| \leq
 K(t, h, \varepsilon, \delta)
 \Vert \nabla T_Q V _{\theta}\Vert \
 \delta ^{{\rm dist } (Q , Q') + {\rm diam} (Q' \cup \pi _{\Sigma } (Q'))
 }$$ which gives us:
$$\sum _{ \lambda \in \Lambda , Q \subseteq \Lambda } |{\rm  Cov }
_{\Lambda , t, \theta} (T_{ \{ \lambda \} } V _{\Lambda , \theta},
h_{Q'})| \leq K(t, h, \varepsilon,  \delta) |\Lambda _{\bot}|.$$
The Proposition is proved, and Theorem 1.5 follows by the same
arguments as Sj\"ostrand [23], Section 8, p.45-46.

\vskip 3cm
 \centerline {\bf References.}
\bigskip

\noindent [1] S. ALBEVERIO,  Y. KONDRATIEV, T. PASUREK, M.
R\"OCKNER, Euclidean Gibbs states of quantum crystals. {\it Moscow
Math. Journal.} {\bf 1}, No 3, (2001), p. 307-313.
 \medskip
\noindent [2] L. AMOUR, M. BEN-ARTZI,  Global existence and decay
for viscous Hamilton-Jacobi equations, {\it Nonlinear Analysis:
Theory, Methods and Applications}, {\bf 31}, 5-6, (1998), 621-628.

\medskip
\noindent [3]L. AMOUR, C. CANCELIER, P. LEVY-BRUHL and J.
NOURRIGAT, {\it  Thermodynamic limits for a quantum crystal by
heat kernel methods. } Universit\'e de Reims, 2003, and mp-arc
03.541.
\medskip \noindent [4]L. AMOUR, C. CANCELIER, P.
LEVY-BRUHL and J. NOURRIGAT, States of a one dimensional quantum
crystal. {\it C. R. Math. Acad. Sci. Paris},  336 (2003), no. 12,
981-984.
\medskip \noindent [5] N. ASHCROFT, D. MERMIN, {\it Solid
State Physics.} Saunders College . Fort Worth, 1976.
\medskip
\noindent [6] V. BACH, J.S.  M\"OLLER, Correlation at low
temperature. I. Exponential decay. {\it J. Funct. Anal.},  {\bf
203} (2003), no. 1, 93-148.
\medskip
\noindent [7] V. BACH, T.
JECKO, J. SJOSTRAND, Correlation asymptotics of classical lattice
spin systems with nonconvex Hamilton function at low temperature.
{\it Ann. Henri Poincar\'e}, {\bf 1}, (2000), no 1, 59-100.

\medskip \noindent
 [8]L. BERTINI,
E.N.M.  CIRILLO et E. OLIVIERI, A combinatorial proof of tree
decay of semi-invariants. {\it J. Statist. Phys. } {\bf 115}
(2004), 395-413.
\medskip \noindent [9] O. BRATTELI, D.W.
ROBINSON,
 {\it Operator algebras and quantum statistical mechanics.
 2. Equilibrium states. Models in quantum statistical mechanics.}
 Second edition. Texts and Monographs in Physics. Springer-Verlag, Berlin, 1997.
 \medskip
\noindent [10] L. GROSS, Decay of correlations in classical
lattice models at high temperature. {\it Comm. in Math. Phys},
{\bf 68} (1979), 1, 9-27.
\medskip \noindent [11]  B. HELFFER,
{\it Semiclassical analysis, Witten Laplacians, and statistical
mechanics.} Series on Partial Differential Equations and
Applications, 1. World Scientific Publishing Co., Inc., River
Edge, NJ, 2002.
\medskip \noindent [12]  B. HELFFER, Remarks on
the decay of correlations and Witten Laplacians, Brascamp-Lieb
inequalities and semi-classical limit, {\it J. Funct. Analysis,}
{\bf 155}, (2), (1998), p.571-586.
\medskip \noindent [13]  B.
HELFFER,  Remarks on the decay of correlations and Witten
Laplacians, II. Analysis of the dependence of the interaction.
{\it Rev. Math. Phys.} {\bf 11} (3), (1999), p.321-336.
 \medskip
\noindent [14]  B. HELFFER,  Remarks on the decay of correlations
and Witten Laplacians, III. Applications to the logarithmic
Sobolev inequalities. {\it Ann. I.H.P. Proba. Stat,} {\bf 35},
(4), (1999), p.483-508.
\medskip \noindent [15] B. HELFFER, J.
SJ\"OSTRAND, On the correlation for Kac like models in the convex
case, {\it J. Stat. Physics},  {\bf 74} (1, 2), (1994), p.349-409.
\medskip \noindent [16] O. MATTE, Supersymmetric Dirichlet
operators, spectral gaps,
  and correlations.
 {\it Ann. Henri Poincar\'e},  {\bf 7} (2006), no. 4, 731-780.
\medskip \noindent
 [17] O. MATTE, J.S. MOLLER, On the spectrum of semi-classical Witten-Laplacians
 and Schr\"odinger  operators in large dimension.
 {\it J. Funct. Anal.} {\bf  220}  (2005), no. 2, 243-264.

\medskip \noindent
 [18]  R. A. MINLOS, {\it Introduction to Mathematical
 Statistical Physics.} University Lecture Series {\bf 19},
 American Mathematical Society, Providence, 2000.
 \medskip
\noindent
 [19] R. A. MINLOS, E.A. PECHERSKY, V. A. ZAGREBNOV,
Analyticity of the Gibbs states  for a quantum anharmonic crystal:
no order parameter. {\it Ann. Henri Poincar\'e} {\bf 3} (2002), p.
921-938.
\medskip \noindent [20] Ch. ROYER, Formes quadratiques et
calcul pseudodiff\'erentiel en grande dimension. {\it
Pr\'e\-publication 00.05.} Reims, 2000.
 \medskip
\noindent [21] D. RUELLE, {\it Statistical Mechanics: rigorous
results.} Addison-Wesley, 1969.
\medskip \noindent [22] B. SIMON,
{\it The statistical Mechanics of lattice gases. } Vol. I.
Princeton Series in Physics. Princeton, 1993.
\medskip \noindent
[23] J. SJ\"OSTRAND, Evolution equations in a large number of
variables, {\it Math. Nachr.} {\bf 166} (1994), 17-53.
\medskip
\noindent [24] J. SJ\"OSTRAND, Correlation asymptotics and Witten
Laplacians, {\it Algebra i Analiz}, {\bf 8} (1996), 1, 160-191.
Translation in {\it St Petersburg Math. Journal}, {\bf 8} (1997),
1, 123-147.
\medskip \noindent [25] J. SJ\"OSTRAND, {\it Complete
asymptotics for correlations of Laplace integrals in the
semiclassical limit.} Memoires S.M.F., {\bf 83}, (2000).

\bigskip
\hskip 9cm laurent.amour@univ-reims.fr
\medskip
\hskip 9cm claudy.cancelier@univ-reims.fr
\medskip
\hskip 9cm  pierre.levy-bruhl@univ-reims.fr
\medskip
\hskip 9cm jean.nourrigat@univ-reims.fr

\end